\documentclass[journal]{IEEEtran}

\IEEEoverridecommandlockouts

\usepackage[cmex10]{amsmath}
\usepackage{amssymb}
\usepackage{graphicx}
\usepackage{subfigure}
\usepackage{cite}
\usepackage{epstopdf}
\usepackage{subeqnarray}
\usepackage{diagbox}
\usepackage{booktabs}
\usepackage{multirow}
\usepackage{multicol}
\usepackage{algorithm}
\usepackage{algorithmic}
\usepackage{url}
\usepackage{xcolor}
\usepackage{array}
\usepackage{footnote}
\usepackage{enumerate}
\usepackage{framed}
\usepackage{lineno}
\usepackage{threeparttable}
\makesavenoteenv{tabular}
\usepackage{amsthm}

\newcommand{\mb}{\mathbf}
\newcommand{\mbb}{\mathbb}

\newcommand{\mr}{\mathrm}
\newcommand{\ti}{\textit}

\newcommand{\tb}{\textbf}

\newtheorem{theorem}{\textbf{Theorem}}
\newtheorem{lemma}{\textbf{Lemma}}

\newtheorem{definition}{\textbf{Definition}}
\newtheorem{example}{\textbf{Example}}

\begin{document}

\title{Basis-Finding Algorithm for Decoding Fountain Codes for DNA-Based Data Storage}

\author{%
\IEEEauthorblockN{Xuan He and Kui Cai}
\thanks{Part of this work has been published in 2020 ITW.}
\thanks{
Xuan He is with the School of Information Science and
Technology, Southwest Jiaotong University, Chengdu 611756, China (e-mail:
xhe@swjtu.edu.cn).}
\thanks{
Kui Cai is with the Science, Mathematics and
Technology (SMT) Cluster, Singapore University of Technology and Design,
Singapore 487372 (e-mail: cai\_kui@sutd.edu.sg).}
}

\maketitle

\begin{abstract}
In this paper, we consider the decoding of fountain codes where the received symbols may have errors. It is motivated by the application of fountain codes in DNA-based data storage systems where the inner code decoding, which generally has undetectable errors, is performed before the outer fountain code decoding. We propose a novel and efficient decoding algorithm, namely basis-finding algorithm (BFA), followed by three implementations. The key idea of the BFA is to find a basis of the received symbols, and then use the most reliable basis elements to recover the source symbols with the inactivation decoding. Gaussian elimination is used to find the basis and to identify the most reliable basis elements. As a result, the BFA has polynomial time complexity. For random fountain codes, we are able to derive some theoretical bounds for the frame error rate (FER) of the BFA. Extensive simulations with Luby transform (LT) codes show that, the BFA has significantly lower FER than the belief propagation (BP) algorithm except for an extremely large amount of received symbols, and the FER of the BFA generally decreases as the average weight of basis elements increases.

\end{abstract}

\begin{IEEEkeywords}
Basis-finding algorithm (BFA), DNA-based data storage, erroneous received symbols, fountain codes, Gaussian elimination.
\end{IEEEkeywords}

\IEEEpeerreviewmaketitle


\section{Introduction}\label{section: introduction}

Fountain codes \cite{byers1998digital, mackay2005fountain} are a class of rateless erasure codes.
They allow to generate and transmit a potentially limitless stream of encoded symbols.
Luby transform (LT) codes \cite{luby2002lt} are the first practical realization of  fountain codes that have fast encoding and decoding algorithms.
Another  well-known type of fountain codes, the Raptor codes \cite{shokrollahi2006raptor}, improve the LT codes by adding a pre-code  such that a linear encoding/decoding complexity (with respect to the number of source symbols) and better error-correction performance can be achieved.

Fountain codes were originally proposed for erasure channels, where an encoded symbol transmitted over the channel is either lost or received without errors \cite{luby2002lt, shokrollahi2006raptor}.
In this case, there exists a fast maximum likelihood (ML) decoder, which is widely referred to  as structured Gaussian elimination (SGE)\cite{odlyzko1984discrete, lamacchia1991solving, he2020disjoint} or inactivation decoding \cite{lazaro2017fountain,  3GPP06, shokrollahi2005systems, shokrollahi2011raptor}.
Later, some works \cite{palanki2004rateless, etesami2006raptor, venkiah2009jointly, ma2006fountain, castura2006rateless} applied fountain codes to other noisy channels, including binary symmetric channel (BSC), additive white Gaussian noise (AWGN) channel, and fading channel.
These works \cite{palanki2004rateless, etesami2006raptor, venkiah2009jointly, ma2006fountain, castura2006rateless} mainly focused on the design of fountain codes and their belief propagation (BP) decoding by using the soft information of the channel.

\begin{figure}[t]
\centering
\includegraphics[scale = 0.5]{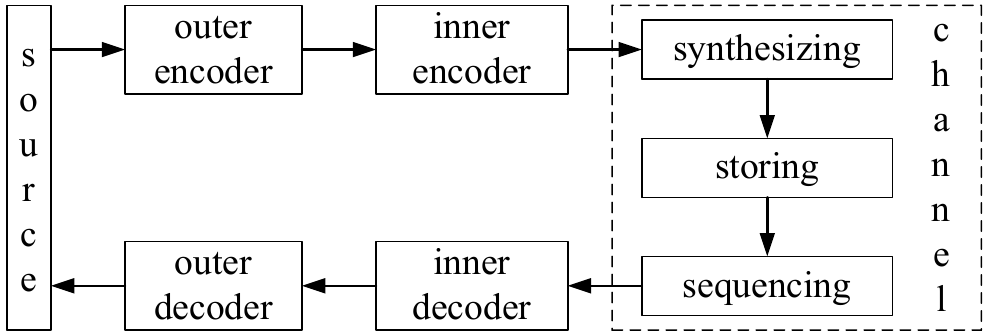}
\caption{
A typical system model for DNA-based data storage.
}
\label{fig: DNA model}
\end{figure}

During recent few years, DNA-based data storage systems attract a lot of attention \cite{organick2018random, nguyen2021capacity, cai2021correcting, liu2022Capacity, 	erlich2017dna}.
Their longevity and extremely high information density make them a promising candidate for archiving massive data in the future.
Fig. \ref{fig: DNA model} shows a typical system model for DNA-based data storage.
The codewords of the inner code are first synthesized (written) into DNA strands, which can be considered as 4-ary ($\{\text{A, T, C, G}\}$) data strings.
The DNA strands are duplicated into many copies before being stored in a DNA pool in an unordered manner.
When needed, the DNA strands are sequenced (read out) in a random sampling fashion from the DNA pool to recover the original information.
The channel model for the aforementioned process is quite complicated. Various types of errors, such as   insertions, deletions, and substitution errors can occur within each DNA strand, and a DNA strand may be totally lost during reading out.
Correspondingly, the inner codes are mainly used to correct or detect the errors within a DNA stand, while the outer codes are mainly used to recover the missing DNA stands and to correct the undetectable errors of the inner codes.

Note that the error rate of DNA strands during the DNA synthesis, storage, and sequencing processes depends heavily on two factors \cite{organick2018random, nguyen2021capacity, cai2021correcting, liu2022Capacity, 	erlich2017dna}: the GC-content which refers to the fraction of nucleotides G and C in a DNA strand, and the homopolymer run which refers to any sub-string of identical nucleotides.
Too high (low) GC-content and/or too long ($> 6$) homopolymer run will dramatically increase the error rate.
Thus, an efficient DNA-based data storage system needs to encode DNA strands (e.g., by constrained coding) with proper GC-content and homopolymer runlength constraints.
However, the design of high rate and low complexity constrained codes that satisfy both the GC-content and homopolymer runlength constraints is a challenge \cite{	nguyen2021capacity, liu2022Capacity}.

In \cite{	erlich2017dna}, an efficient DNA-based data storage architecture named DNA fountain was proposed, which featured  a fountain code for a set of DNA strands as the outer code.
It can effectively overcome the missing of DNA strands.
Note that in this case, each DNA strand corresponds to an encoded symbol of the outer fountain code.
We would like to highlight that, compared to other error-correction codes such as the low-density parity-check (LDPC) codes \cite{Gallager62} and polar codes \cite{arikan2009channel}, adopting the fountain codes as the outer codes of DNA-based data storage systems has an obvious advantage: no specifically designed constrained codes are needed to impose the GC-content constraint and the homopolymer runlength constraint to the DNA strand, since the fountain codes can potentially generate a limitless number of encoded symbols due to the rateless property.
Hence the encoded symbols that violate the GC-content and homopolymer runlength constraints  can simply be discarded until the sufficient number of encoded symbols that satisfy the required constraints  are generated.
Mainly due to this advantage, the DNA fountain scheme can achieve the highest net information density among various other  DNA-based data storage architectures \cite{erlich2017dna}.

On the other hand, in the DNA fountain, a Reed-Solomon (RS) code \cite{ECC04} is adopted for each DNA strand as the inner code, which can detect or correct the errors within a DNA strand.
However, to ensure a high net information density, only a limited amount of redundancy can be introduced to the RS code, which may not be sufficient to detect or correct all insertions, deletions, and substitution errors that may occur within a DNA strand.
As a result, there is a non-negligible chance to have undetectable errors within the DNA strands at the output of the RS decoder \cite{	erlich2017dna}.
Hence, these erroneous DNA strands become a killing factor for the decoding of the outer fountain code with the inactivation decoding \cite{odlyzko1984discrete, lamacchia1991solving, he2020disjoint, lazaro2017fountain,  3GPP06, shokrollahi2005systems, shokrollahi2011raptor}.

Motivated by the application of fountain codes in DNA-based data storage systems and the corresponding problems encountered as described above,  we consider the decoding of fountain codes where each received symbol may have errors.
That is, each received symbol is correct with probability $p$, and is corrupted by substitution errors with probability $1-p$.
Note that for a fixed symbol size, the substitution errors within a received symbol may also be caused by  equal number of insertions and deletions.
We remark that in the DNA fountain scheme \cite{	erlich2017dna}, since the inner code decoding is performed before the outer fountain code decoding, a received symbol corresponds to a decoded codeword of the inner code and it generally has a high probability to be correct (i.e., $p$ is close to 1).
Given the channel transition probability, the BP decoder \cite{etesami2006raptor} is applicable for decoding, but the inactivation decoding \cite{odlyzko1984discrete, lamacchia1991solving, he2020disjoint, lazaro2017fountain,  3GPP06, shokrollahi2005systems, shokrollahi2011raptor} is not due to the existence of erroneous received symbols.

We propose a novel and efficient decoding algorithm, namely basis-finding algorithm (BFA).
The key idea of the BFA is to first find a basis of the received symbols, and then use the most reliable basis elements to recover the source symbols with the inactivation decoding (same as the decoding of fountain codes for erasure channels).
For the case with $p = 1$ which corresponds to an erasure channel, the BFA is essentially the same as the inactivation decoding \cite{odlyzko1984discrete, lamacchia1991solving, he2020disjoint, lazaro2017fountain,  3GPP06, shokrollahi2005systems, shokrollahi2011raptor}.
From this point of view, the BFA can be regarded as a generalization of the inactivation decoding for $p < 1$.

Naturally, Gaussian elimination is used to  find the basis and to identify the most reliable basis elements, leading to a straightforward implementation for the BFA.
Through extensive simulations, we observe that:
\begin{itemize}
    \item   \tb{Observation 1:} A basis with larger average weight of its basis elements generally has better error-correction performance.
\end{itemize}
Motivated by the above observation, we optimize the average weight of basis elements by introducing a pre-processing step to rearrange the received symbols before applying the straightforward implementation.
Such an operation is incorporated into two new implementations for the BFA, namely sorted-weight implementation and triangulation-based implementation.

Using random fountain codes, we derive some theoretical bounds for the frame error rate (FER) of the BFA.
Moreover,  we perform extensive simulations using the considered LT codes to evaluate the FER of the BFA.
The simulation results show that for the considered channel, the BFA (with any of the three implementations) has significantly lower FER than the BP algorithm \cite{etesami2006raptor} except for $p < 1$ and an extremely large number of received symbols.
It is also shown that the straightforward implementation, triangulation-based implementation, and  sorted-weight implementation has higher to lower FERs and smaller to larger average weights of basis elements.
It is worth mentioning that the BFA only has simple integer operations, many of which are essentially bit-XOR operations; on the contrary, the BP algorithm has many complicated floating-point operations.

The remainder of this paper is organized as follows.
Section \ref{section: preliminary} introduces the LT codes, which are the first practical realization of fountain codes and will be used in our simulations.
Section \ref{section: system model} illustrates the system model considered by this work.
Section \ref{section: Basis-Finding Algorithm} proposes a general framework of the BFA.
Section \ref{section: implementation} develops three implementations for the BFA, namely the straightforward implementation, the sorted-weight implementation, and the triangulation-based implementation.
We derive theoretical bounds for the FER of the BFA in Section \ref{section: performance analysis}, and present the simulation results in Section \ref{section: simulation}.
Finally, we conclude the paper in Section \ref{section: conclusion}.

\ti{Notations:} In this paper, we use non-bold small letters for scalars (e.g., $m$), bold small letters for vectors (e.g., $\mb{a}$), and bold capital letters for matrices (e.g., $\mb{A}$).
For any positive integer $m$, define $[m] := \{1, 2, \ldots, m\}$ and denote $0^m$ as a zero vector of length $m$.
Denote $\mbb{F}_2$ as the binary field and $\mbb{R}$ as the real domain.
Let $\mbb{P}(\cdot)$ denote the probability of an event and $\mr{rank}(\cdot)$ denote the rank of a matrix.

\section{Preliminaries}\label{section: preliminary}

We consider to use LT codes \cite{	luby2002lt} in this paper.
However, our proposed algorithm is also applicable to Raptor codes \cite{	shokrollahi2006raptor}.
Suppose that there are $n$ source symbols denoted by
\[
    \mb{X} :=
    \left[
        \begin{array}{c}
            \mb{x}_1\\
            \mb{x}_2\\
            \vdots\\
            \mb{x}_n\\
        \end{array}
    \right]
    \in \mbb{F}_2^{n \times l},
\]
where $\mb{x}_i, i \in [n]$ denotes the $i$-th source symbol which is a bit string of length $l$.
The sender can generate and send potentially a limitless number of  encoded symbols.
We denote an encoded symbol by a two-tuple
\[
    (\mb{a}, \mb{y}) \in (\mbb{F}_2^n, \mbb{F}_2^l), \text{~s.t.~} \mb{y} = \mb{a} \mb{X} = \oplus_{i \in [n]: a_i = 1} \mb{x}_i,
\]
where $\oplus$ is the bitwise addition over $\mbb{F}_2$ and $a_i$ is the $i$-th entry of $\mb{a}$.
We call $\mb{a}$ and $\mb{y}$ as the constraint and data payload of the encoded symbol $(\mb{a}, \mb{y})$, respectively.

For the encoding process, we only need to define a distribution for selecting the weight of an encoded symbol (i.e., the number of ones in $\mb{a}$), say $d$, and set the $d$ ones of $\mb{a}$ uniformly at random.
The well-known ideal soliton distribution (ISD)  $\rho(\cdot)$ \cite{luby2002lt} and robust soliton distribution (RSD) $\mu(\cdot)$  \cite{luby2002lt} for selecting $d$ are defined below.

\begin{definition}[ISD and RSD \cite{luby2002lt}]\label{def: ISD/RSD}
    The ISD  $\rho(\cdot)$ is defined by
    \begin{equation*}
        \rho(d) =
        \begin{cases}
            \frac{1}{n}, & d = 1,\\
            \frac{1}{d (d-1)},  & d = 2, 3, \dots, n.
        \end{cases}
    \end{equation*}
    For suitable constants $\delta, c > 0$, let $R = c \sqrt{n} \ln(n/\delta)$.
    Define a function $\tau: [n] \to \mbb{R}$ by
    \begin{equation*}
        \tau(d) =
        \begin{cases}
            \frac{R}{d n}, & d = 1, 2, \dots, \frac{n}{R} - 1,\\
            \frac{R}{n} \ln\frac{R}{\delta}, & d = \frac{n}{R},\\
            0, & \text{otherwise.}
        \end{cases}
    \end{equation*}
    The RSD is defined by
    \[
        \mu(d) = \frac{\rho(d) + \tau(d)}{\beta}, d \in [n],
    \]
    where $\beta = \sum_{d \in [n]} \rho(d) + \tau(d)$.
\end{definition}

In practice, not $\mb{a}$, but a corresponding seed using which a predefined pseudo-random number generator (PRNG)  can generate $\mb{a}$, is transmitted over the channel so as to save resources.
Suppose the number of different transmitted encoded symbols is $m$.
In general, the size of $m$ depends on the channel and we have $n \leq m \ll 2^n$, where $2^n$ is the maximum possible number of different encoded symbols generated from $n$ source symbols.
Then, a seed needs at least $\log_2(m) \ll n$ bits.
As a result, the code rate is upper bounded by $n l / (m l +  m \log_2(m)) = n / (m  (1 + \log_2(m) / l))$.
We can see that smaller  $\log_2(m) / l$ means  more negligible cost for using seed as well as  higher transmission efficiency.
Therefore, to ensure a high transmission efficiency, it is generally required that  $\log_2(m) \ll l$.

Suppose that the receiver collects $m$ encoded symbols which are denoted by
\[
    (\mb{A}, \mb{Y}) :=
    \left[
        \begin{array}{c c}
            \mb{a}_1 & \mb{y}_1\\
            \mb{a}_2 & \mb{y}_2\\
            \vdots & \vdots\\
            \mb{a}_m & \mb{y}_m\\
        \end{array}
    \right]
    \in (\mbb{F}_2^{m \times n}, \mbb{F}_2^{m \times l}),
\]
where the $i$-th row $(\mb{a}_i, \mb{y}_i) \in \mbb{F}_2^{n+l}, i \in [m]$ denotes the $i$-th received symbol.
For erasure channels, all received symbols are correct, i.e., $\mb{A} \mb{X} = \mb{Y}$.
Under such a situation, the inactivation decoding \cite{odlyzko1984discrete, lamacchia1991solving, he2020disjoint, lazaro2017fountain,  3GPP06, shokrollahi2005systems, shokrollahi2011raptor} can efficiently recover $\mb{X}$ from $(\mb{A}, \mb{Y})$ if $\mr{rank}(\mb{A}) = n$.
In general, $\mr{rank}(\mb{A}) = n$ is easily satisfied when the overhead, defined by $m - n$, is positive and not too small.
For random fountain codes, i.e., $\mb{A}$ is independently and uniformly chosen from $\mbb{F}_2^{m \times n}$, we have the following result.

\begin{lemma}\label{lemma: rank property}
Assume $\mb{A}$ is independently and uniformly chosen from $\mbb{F}_2^{m \times n}$ with $m \geq n \geq 0$.
The probability for $\mr{rank}(\mb{A}) = n$  is given by
\[
    P_{rk}(m, n) = \prod_{i = 0}^{i = n - 1} (1 - 2^{i - m}) \geq 1 - 2^{n - m},
\]
where for $n = 0$, we have $P_{rk}(m, n) = 1$.
\end{lemma}

\begin{IEEEproof}
For $n = 0$, it is obvious to have $P_{rk}(m, n) = 1 \geq 1 - 2^{- m}$.
For $n > 0$, each column of $\mb{A}$ is independently and uniformly chosen from $\mbb{F}_2^{m}$ by assumption.
Let $p_i, i \in [n]$ denote the probability that the first $i$ columns are linearly independent.
We have $p_1 = 1 - 2^{-m}$ and $p_{i + 1} = p_{i} (1 - 2^{i - m}), \forall i \in [n - 1]$.
Thus, we have $P_{rk}(m, n) = p_n = \prod_{i = 0}^{i = n - 1} (1 - 2^{i - m}) \geq 1 - \sum_{i = 0}^{i = n - 1} 2^{i - m} \geq 1 - 2^{n - m}$.
\end{IEEEproof}

Lemma \ref{lemma: rank property} indicates that for  an arbitrary matrix uniformly taken from $\mbb{F}_2^{m \times n}$ with $m \geq n$, this matrix has a high probability to have $n$ linearly independent rows.
This also implies that random fountain codes have excellent error-correction performance (at the cost of high encoding and decoding complexities).
We remark that good Raptor codes add dense rows to mimic the behavior of random fountain codes \cite{	shokrollahi2011raptor}.
Moreover, we present another similar lemma below.
Both Lemmas \ref{lemma: rank property} and \ref{lemma: rank property no zero row} are important for the error-correction performance analysis in Section \ref{section: performance analysis}.

\begin{lemma}\label{lemma: rank property no zero row}
Assume $\mb{A}$ is independently and uniformly chosen from $\mbb{F}_2^{m \times n}$ with $m \geq n \geq 0$.
In addition, each column of $\mb{A}$ is not a zero vector.
The probability for $\mr{rank}(\mb{A}) = n$  is given by
\[
    P_{rk}^*(m, n) = \prod_{i = 0}^{i = n - 1} \left(1 - \frac{2^{i} - 1}{2^{m} - 1}\right) \geq 1 - 2^{n - m},
\]
where for $n = 0$, we have $P_{rk}^*(m, n) = 1$.
\end{lemma}

\begin{IEEEproof}
For $n = 0$, it is obvious to have $P_{rk}^*(m, n) = 1 \geq 1 - 2^{- m}$.
For $n > 0$, each column of $\mb{A}$ is independently and uniformly chosen from $\mbb{F}_2^{m} \setminus \{0^m\}$ by assumption.
Let $p_i, i \in [n]$ denote the probability that the first $i$ columns are linearly independent.
We have $p_1 = 1$ and $p_{i + 1} = p_{i} (1 - \frac{2^{i} - 1}{2^{m} - 1}), \forall i \in [n - 1]$.
Thus, we have $P_{rk}^*(m, n) = p_n = \prod_{i = 0}^{i = n - 1} (1 - \frac{2^{i} - 1}{2^{m} - 1}) \geq \prod_{i = 0}^{i = n - 1} (1 - 2^{i - m}) \geq 1 - 2^{n - m}$.
\end{IEEEproof}

We remark that for any $0 < n \leq m$, we have
\begin{align*}
    P_{rk}^*(m, n) &= P_{rk}(m, n)\frac{2^m}{2^m - 1}
    \\&= \frac{P_{rk}(m, m)}{P_{rk}(m-n, m-n)} \frac{2^m}{2^m - 1}.
\end{align*}
In addition, for any $i > 0$, we have
\[
    P_{rk}(i, i) = P_{rk}(i-1, i-1) (1 - 2^{-i}).
\]
This indicates that we can precompute and store $P_{rk}(i, i)$ for all $i \in [m]$ with time complexity $O(m)$ and storage complexity $O(m)$, respectively.
We can then  compute $P_{rk}(i, j)$ and/or $P_{rk}^*(i, j)$ on-the-fly with time complexity $O(1)$ for any $m \geq i \geq j > 0$.

\section{System Model}\label{section: system model}

\begin{figure}[!t]
\centering

\subfigure[A simplified system model for DNA-based data storage. For channel-1, rows of $\mb{Z}$ are transmitted and received one-by-one in order, where each row is received correctly with probability $p$ and corrupted by substitution errors with probability $1 - p$.
For channel-2, $\mb{Y}'$ is a random permutation of the rows of $\mb{Y}$.]{%
    \includegraphics[scale = 0.5]{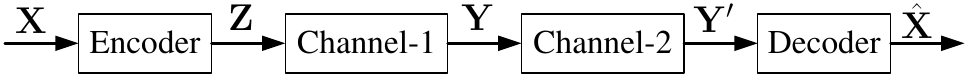}
}

\subfigure[A simplified system model for DNA-based data storage when applying fountain codes as the outer codes. Rows of $(\mb{A}, \mb{Z})$ are transmitted and received one-by-one in order, where $\mb{A}$ is always received correctly while  each row of $\mb{Z}$ is received correctly with probability $p$ and corrupted by substitution errors with probability $1 - p$.]{%
    \includegraphics[scale = 0.5]{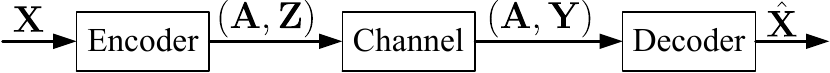}
}

\caption{System model considered in this paper..}
\label{fig: channel}
\end{figure}

In this paper, we are interested in the decoding of the outer fountain codes with respect to the system model shown in Fig. \ref{fig: DNA model}.
Therefore, we consider the inner code as part of the channel such that the channel outputs are the valid  codewords of the inner code (may have undetectable errors) generated by its decoder.
This leads to the simplified channel models illustrated by Fig. \ref{fig: channel}.

Consider the generic simplified  model shown in Fig. \ref{fig: channel}(a) which does not necessarily use the fountain codes.
The source information $\mb{X}$ is first encoded into $\mb{Z} \in \mbb{F}_2^{m \times w}$.
Then, $\mb{Z}$ is transmitted over the channel and $\mb{Y}'$ is the final  channel output.
Hence, the channel is the concatenation of two  sub-channels, i.e., channel-1 and channel-2.
For channel-1, rows of $\mb{Z}$ are transmitted and received one-by-one in order.
Since each row indeed corresponds to a codeword of the inner codes, we do not model channel-1 by independently introducing noise to each transmitted bit (like the modeling of BSC).
Instead, we consider the situation that each row is transmitted per channel use, and the row is received correctly with probability $p$ and corrupted by substitution errors with probability $1 - p$, where $p$ is a predefined and fixed positive constant throughout this paper.
Moreover, it is reasonable to think that, each row is correct with high probability (i.e., $p$ is large) and when it is not correct, it generally suffers from random substitution errors\footnote{For example, when insertions and deletions occur to a row and it becomes a valid codeword with undetectable substitution errors after the inner code decoding, the substitution errors are likely to occur randomly among the bits within a row.}.
For channel-2, it permutates the order of the rows of $\mb{Y}$ uniformly at random, corresponding to the unordered manner in DNA-based data storage systems.

Note that the channel in Fig. \ref{fig: channel}(a) (by regarding each transmitted row as a single symbol) is identical to the noisy permutation channel in \cite[Fig. 2]{makur2020coding}, where the channel capacity  is considered for the case that $w > 0$ is fixed and $m \to \infty$.\footnote{The definition for the channel capacity in \cite{makur2020coding} is  different from the conventional one. In fact, the conventional definition leads to zero capacity in this case according to \cite{Shomorony2021DNA}.}
Meanwhile, the channel in Fig. \ref{fig: channel}(a) with channel-1 being replaced by a BSC or an erasure channel (each row is either lost or received correctly)  is a special case of the noisy shuffling-sampling channel  in \cite{Shomorony2021DNA}.
It was proved in \cite{Shomorony2021DNA} that  for  $m \to \infty$ and a fixed $\beta := \lim_{m \to \infty} w / \log_2(m) \leq 1$, the capacity of the noisy shuffling-sampling channel  is 0; for  $m \to \infty$ and a fixed $\beta > 1$, appending a unique index of bit-width $\log_2(m)$ to each row of $\mb{Z}$ does not incur rate loss (i.e., the index-based coding scheme is optimal).
However, the capacity of the channel in Fig. \ref{fig: channel}(a) is generally not known (e.g., for $0 < p < 1$ and $\beta > 1$) up till now.
This problem may be solved based on the techniques used in \cite{makur2020coding} and/or \cite{Shomorony2021DNA}.
Here we provide  a simple estimation of the capacity.
Assume that there is a genie who helps to identify and discard the incorrect rows received from channel-1.
As a result, channel-1 becomes an erasure channel and the overall capacity of the modified channel in Fig.  \ref{fig: channel}(a)  is given by
\cite{Shomorony2021DNA}
\begin{align}\label{eqn: c erasure}
    C_{\text{erasure}} \sim p(1 - 1/\beta)w = p(w - \log_2(m)), \text{~s.t.~} \beta > 1,
\end{align}
in which $- \log_2(m)$ can be considered as the loss due to the random permutation of channel-2.
($C_{\text{erasure}} \neq p(1 - 1/\beta)$ since a row of $w$ bits is transmitted per channel use.)
Since the discarded incorrect rows generally  suffer from random substitution errors and thus can hardly provide any information about the transmitted rows, $C_{\text{erasure}}$ turns to be a reasonable upper bound for the capacity of the channel in Fig. \ref{fig: channel}(a).
In fact, later in Section \ref{section: performance analysis}, we will show that $C_{\text{erasure}}$ is achievable in some cases by the fountain codes under the proposed decoding algorithm (i.e., BFA).

We now turn to the system model illustrated by Fig. \ref{fig: channel}(b), which is obtained by applying fountain codes to the system model in Fig. \ref{fig: channel}(a).
At the sender side, the $n$ source symbols $\mb{X} \in \mbb{F}_2^{n \times l}$ are encoded to  $m$ encoded symbols $(\mb{A}, \mb{Z}) \in (\mbb{F}_2^{m \times n}, \mbb{F}_2^{m \times l})$, where $\mb{A} \mb{X} = \mb{Z}$.
Then, the rows (encoded symbols) of $(\mb{A}, \mb{Z})$ are transmitted and received one-by-one in order, where $\mb{A}$ is always received correctly while  each row of $\mb{Z}$ is received correctly with probability $p$ and corrupted by substitution errors with probability $1 - p$.
Finally, the received symbols $(\mb{A}, \mb{Y})$ are used to perform decoding and to provide an estimation $\hat{\mb{X}}$ of $\mb{X}$.

Here we show the rationale  of the system model in  Fig. \ref{fig: channel}(b).
Given the source symbols $\mb{X}$, an encoded symbol $(\mb{a}, \mb{y})$ is correct if and only if (iff) $\mb{a} \mb{X} =  \mb{y}$.
We can consider $\mb{a}$ as the index of $(\mb{a}, \mb{y})$, and regard fountain codes as an index-based coding scheme.
Given the fact that encoded symbols $(\mb{A}, \mb{Y})$  are received, the decoding performance relies on the number of correct received symbols in $(\mb{A}, \mb{Y})$ rather than the order of the received symbols.
Therefore, we can reorder the received symbols if needed.
Hence, it is reasonable to assume that the encoded symbols are transmitted and received one-by-one in order in the system model in Fig. \ref{fig: channel}(b) for the sake of convenience.
On the other hand, suppose $(\mb{a}, \mb{y})$ is sent and $(\mb{a}', \mb{y}')$ is received, and some errors occur in the seed corresponding to $\mb{a}$, leading to $\mb{a} \neq \mb{a}'$.
However, according to the encoding process introduced in Section \ref{section: preliminary}, $\mb{a}$ and $\mb{a}'$ must follow the same weight distribution, and thus $(\mb{a}', \mb{a}' \mb{X})$ works similarly to $(\mb{a}, \mb{y})$  statistically in the decoding of fountain codes.
As a result, we can regard $(\mb{a}', \mb{a}' \mb{X})$ as the original transmitted encoded symbol leading to $(\mb{a}', \mb{y}')$.
Therefore, it is reasonable to assume that errors only occur in $\mb{Y}$ but not in $\mb{A}$ in the system model in Fig. \ref{fig: channel}(b).
Moreover, when errors occur in a seed, it is reasonable to consider that it is  equivalent to the case where uniformly random errors occur in data payload.

According to the above discussion, we define the error pattern of $(\mb{A}, \mb{Y})$ by
\begin{equation}\label{eqn: error pattern}
    \mb{S} :=
    \left[
        \begin{array}{c}
            \mb{s}_1\\
            \mb{s}_2\\
            \vdots\\
            \mb{s}_m\\
        \end{array}
    \right]
    =
    \left[
        \begin{array}{c}
            \mb{z}_1 \oplus  \mb{y}_1\\
            \mb{z}_2 \oplus \mb{y}_2\\
            \vdots\\
            \mb{z}_m \oplus \mb{y}_m\\
        \end{array}
    \right]
    \in \mbb{F}_2^{m \times l},
\end{equation}
which is actually not available to the receiver since $\mb{X}$ and $\mb{Z}$ are unknown.
However, the receiver may know that
\[
    \mbb{P}(\mb{s}_i = 0^l) = p, \forall i \in [m].
\]
Since generally $p$ is large, we assume that
\begin{equation}\label{eqn: p max}
      p > \mbb{P}(\mb{s}_i = \mb{s}), ~\forall \mb{s} \in \mbb{F}_2^l \setminus \{0^l\}.
\end{equation}
We may further assume that each incorrect received symbol suffers from uniformly random errors when needed, i.e.,
\begin{equation}\label{eqn: random error}
        \mbb{P}(\mb{s}_i = \mb{s}) = \frac{1 - p}{2^l - 1}, ~\forall \mb{s} \in \mbb{F}_2^l \setminus \{0^l\}.
\end{equation}
Although the assumption of \eqref{eqn: random error} is reasonable for DNA-based data storage systems, the BFA proposed in the next section does not rely on the specific values of $\mbb{P}(\mb{s}_i = \mb{s}), \forall \mb{s} \in \mbb{F}_2^l$ to perform decoding.

\ti{BP decoding:} To end this section, we give some remarks about the BP decoder \cite{etesami2006raptor}.
The BP decoder   relies on the specific values of $\mbb{P}(\mb{s}_i = \mb{s}), \forall \mb{s} \in \mbb{F}_2^l$ to compute soft information, say log-likelihood ratios (LLRs), so as to perform decoding.
For example, under the assumption of \eqref{eqn: random error} and assume that the data payload bits follow uniform distribution, the probability of each data payload bit in a received symbol to be correct is given by
\begin{equation}\label{eqn: pb}
    p_{b} = p + (1 - p) (2^{l-1} - 1)/(2^{l} - 1).
\end{equation}
It can be used to compute the LLR of the bit and to  perform decoding of each bit within a received symbol.
We name this the  bit-level BP decoding.
Meanwhile, it is theoretically possible to consider each received symbol as a whole and perform decoding over symbols, such as the BP decoding of non-binary LDPC codes \cite{davey1998low}.
We call this the symbol-level BP decoding, which can have much better error-correction performance than the bit-level BP decoding.
However, the symbol-level BP decoding suffers from extremely high computational complexity, which is $O(2^{2l})$ for each non-zero entry of $\mb{A}$.
We generally have $l \sim 100$ for the current DNA-based data storage systems \cite{erlich2017dna} and $l$ will further increase in the future.
As a result, the symbol-level BP decoding is not practical for DNA-based data storage systems, and we thus only consider to apply bit-level BP decoding in this paper.

\section{Basis-Finding Algorithm}\label{section: Basis-Finding Algorithm}

In this section, we propose the BFA for decoding fountain codes under the system model illustrated in Section \ref{section: system model}.
Let $\mb{D}$ be an arbitrary matrix over $\mbb{F}_2$ with $\mr{rank}(\mb{D}) > 0$.
With a little abuse of notations, we may also use $\mb{D}$ as a multiset (a set allows for multiple instances for each of its elements) which takes all its rows as the set elements.
We define a basis of $\mb{D}$, denoted by $B(\mb{D})$, as a set that consists of $\mr{rank}(\mb{D})$ linearly independent rows of $\mb{D}$.
Given $B(\mb{D})$, for any row $\mb{d} \in \mb{D}$, there exists a unique subset $B_{\mb{d}} \subseteq B(\mb{D})$ such that $\mb{d} = \oplus_{\mb{b} \in B_{\mb{d}}} \mb{b}$.
We name $\oplus_{\mb{b} \in B_{\mb{d}}} \mb{b}$ the linear combination (LC) of the basis elements in $B(\mb{D})$ that represents $\mb{d}$, or equivalently, name it the linear representation (LR) of $\mb{d}$.
A basis element $\mb{b}' \in B(\mb{D})$ is said to attend $\oplus_{\mb{b} \in B_{\mb{d}}} \mb{b}$ if $\mb{b}' \in B_{\mb{d}}$.
The LRs of $\mb{D}$ refer to all the LRs of the rows of $\mb{D}$.

\begin{example}
For example, assume $(n, l, m) = (2, 2, 5)$ and consider the following matrix formed by five received symbols:
\begin{equation}\label{eqn: eg A, y}
    (\mb{A}, \mb{Y}) =
    \left[
        \begin{array}{c c c c}
            1 & 1 & 0 & 1\\
            1 & 0 & 1 & 1\\
            1 & 1 & 1 & 0\\
            0 & 1 & 0 & 1\\
            1 & 0 & 0 & 0\\
        \end{array}
    \right].
\end{equation}
We have $\mr{rank}(\mb{A}, \mb{Y}) = 3$.
Let $B(\mb{A}, \mb{Y})$ consist of the first three rows of $(\mb{A}, \mb{Y})$.
We can easily verify that $B(\mb{A}, \mb{Y})$ is a basis of $(\mb{A}, \mb{Y})$.
Then, each row of $(\mb{A}, \mb{Y})$ can be uniquely represented as an LC of $B(\mb{A}, \mb{Y})$.
For example, we have
\begin{align*}
(\mb{a}_4, \mb{y}_4) &= (\mb{a}_2, \mb{y}_2) \oplus (\mb{a}_3, \mb{y}_3),\\
(\mb{a}_5, \mb{y}_5) &= (\mb{a}_1, \mb{y}_1) \oplus (\mb{a}_2, \mb{y}_2) \oplus (\mb{a}_3, \mb{y}_3).
\end{align*}
Accordingly, we say that $(\mb{a}_1, \mb{y}_1)$ attends the LR of $(\mb{a}_5, \mb{y}_5)$, and both $(\mb{a}_2, \mb{y}_2)$ and $(\mb{a}_3, \mb{y}_3)$ attend the LRs of $\{(\mb{a}_4, \mb{y}_4), (\mb{a}_5, \mb{y}_5)\}$.
\end{example}

For any source symbols $\mb{X} \in \mbb{F}_2^{n \times l}$, denote
\begin{equation}\label{eqn: V(x)}
    V(\mb{X}) = \{(\mb{a}, \mb{y}) \in (\mbb{F}_2^n, \mbb{F}_2^l): \mb{y} = \mb{a} \mb{X}\}.
\end{equation}
$V(\mb{X})$ contains all the $2^n$ possible encoded symbols/correct received symbols (regardless of the weight distribution), each of which is a row vector in $\mbb{F}_2^{n+l}$.
Any basis of $V(\mb{X})$, say $B(V(\mb{X}))$, must contain exactly $n$ linearly independent encoded symbols.
We can recover $\mb{X}$ from $B(V(\mb{X}))$ by using the inactivation decoding \cite{odlyzko1984discrete, lamacchia1991solving, he2020disjoint, lazaro2017fountain,  3GPP06, shokrollahi2005systems, shokrollahi2011raptor}.
Inspired by the above observations, our proposed BFA consists of three steps:
(i) Find a basis of $(\mb{A}, \mb{Y})$,  denoted by $B(\mb{A}, \mb{Y})$; (ii) Identify the $n$ most reliable basis elements from $B(\mb{A}, \mb{Y})$ according to the number of LRs each basis element attends; (iii) Recover $\mb{X}$ from the $n$ most reliable basis elements with the inactivation decoding \cite{odlyzko1984discrete, lamacchia1991solving, he2020disjoint, lazaro2017fountain,  3GPP06, shokrollahi2005systems, shokrollahi2011raptor}.
We formally describe the outline of BFA in  Algorithm \ref{algo: BFA}.

\begin{algorithm}[t!]
    \caption{Basis-finding algorithm (BFA)}
    \label{algo: BFA}
    \begin{algorithmic}[1]
        \REQUIRE    $(\mb{A}, \mb{Y})$.
        \ENSURE     $\hat{\mb{X}}$.
        \STATE  \ti{Step 1 (Find a basis of $(\mb{A}, \mb{Y})$):} Assume $\mr{rank}(\mb{A}, \mb{Y}) \geq n$, otherwise we can never recover $\mb{X}$.
        Find a basis of $(\mb{A}, \mb{Y})$, denoted by $B(\mb{A}, \mb{Y})$, which consists of $\mr{rank}(\mb{A}, \mb{Y})$ linearly independent rows of $(\mb{A}, \mb{Y})$.

        \STATE  \ti{Step 2 (Find the $n$ most reliable basis elements of $B(\mb{A}, \mb{Y})$):}
        Find the $n$ basis elements of $B(\mb{A}, \mb{Y})$, denoted by the set $B_n(\mb{A}, \mb{Y})$, which attend the largest numbers of linear representations of $(\mb{A}, \mb{Y})  \setminus B(\mb{A}, \mb{Y})$.

        \STATE  \ti{Step 3 (Recover $\mb{X}$):} Give an estimation $\hat{\mb{X}}$ of $\mb{X}$ from $B_n(\mb{A}, \mb{Y})$ by using the inactivation decoding \cite{odlyzko1984discrete, lamacchia1991solving, he2020disjoint, lazaro2017fountain,  3GPP06, shokrollahi2005systems, shokrollahi2011raptor}.
    \end{algorithmic}
\end{algorithm}

In Step 1 of Algorithm \ref{algo: BFA}, we can easily generate $B(\mb{A}, \mb{Y})$ in the way below:
\begin{enumerate}[(W1)]
\item   Let $B(\mb{A}, \mb{Y}) = \emptyset$.
\item   For $i = 1, 2, \ldots, m$, if $(\mb{a}_i, \mb{y}_i)$ is not an LC of $B(\mb{A}, \mb{Y})$, add  $(\mb{a}_i, \mb{y}_i)$ into $B(\mb{A}, \mb{Y})$.
\end{enumerate}
The rationale behind Step 2 of Algorithm \ref{algo: BFA} is that the correct basis elements of $B(\mb{A}, \mb{Y})$ generally attend more LRs of $(\mb{A}, \mb{Y})  \setminus B(\mb{A}, \mb{Y})$ than the incorrect basis elements, which provides a way to measure the reliability of each basis element.
We will discuss this in detail later in Section \ref{section: performance analysis} when analyzing the error-correction performance of Algorithm \ref{algo: BFA}.
Accordingly, $B_n(\mb{A}, \mb{Y})$ has a high chance to have $n$ correct basis elements.
If $B_n(\mb{A}, \mb{Y})$ consists of $n$ correct basis elements, Step 3 of Algorithm \ref{algo: BFA} can successfully recover $\mb{X}$ (i.e., $\hat{\mb{X}} = \mb{X}$) with the inactivation decoding \cite{odlyzko1984discrete, lamacchia1991solving, he2020disjoint, lazaro2017fountain,  3GPP06, shokrollahi2005systems, shokrollahi2011raptor}.

\begin{example}
We give a toy example to show how Algorithm \ref{algo: BFA} works.
Assume that $(n, l, m) = (2, 2, 5)$, $\mb{X} =
\left[
    \begin{smallmatrix}
    1 & 1 \\ 0 & 1
    \end{smallmatrix}
\right]$,
and $(\mb{A}, \mb{Y})$ is given by \eqref{eqn: eg A, y}.
In this case, only $(\mb{a}_1, \mb{y}_1)$ and $(\mb{a}_5, \mb{y}_5)$ are incorrect.
In Step 1, according to (W1) and (W2), we have $B(\mb{A}, \mb{Y}) = \{(\mb{a}_1, \mb{y}_1), (\mb{a}_2, \mb{y}_2), (\mb{a}_3, \mb{y}_3)\}$.
In Step 2, we have $B_n(\mb{A}, \mb{Y}) = \{(\mb{a}_2, \mb{y}_2), (\mb{a}_3, \mb{y}_3)\}$, since both $(\mb{a}_2, \mb{y}_2)$ and $(\mb{a}_3, \mb{y}_3)$ attend two LRs of $(\mb{A}, \mb{Y})  \setminus B(\mb{A}, \mb{Y}) = \{(\mb{a}_4, \mb{y}_4), (\mb{a}_5, \mb{y}_5)\}$, while $(\mb{a}_1, \mb{y}_1)$ only attends one LR (the LR of $(\mb{a}_5, \mb{y}_5)$).
By assumption, $B_n(\mb{A}, \mb{Y})$ consists of two correct received symbols, indicating that Step 3 can successfully recover $\mb{X}$ from $B_n(\mb{A}, \mb{Y})$.
\end{example}

We remark that the BFA is very simple in the sense that it only needs to involve the basic operations of  linear algebra, e.g. Gaussian elimination.
We will show how to implement it in detail in the next section.
We further remark that the BFA only needs to take the received symbols $(\mb{A}, \mb{Y})$ as input.
Other information, such as the specific values of $\mbb{P}(\mb{s}_i = \mb{s}), \forall \mb{s} \in \mbb{F}_2^l$, is not necessary.
However, if we have some reliability information associated with the received symbols, the BFA can make use of it: Process the received symbols from the most reliable one to the least reliable one when generating the basis $B(\mb{A}, \mb{Y})$ in Step 1.
By doing so, the basis $B(\mb{A}, \mb{Y})$ can have the best chance to include $n$ correct basis elements, which is necessary for successfully recovering $\mb{X}$.
For example, in the DNA fountain scheme \cite{	erlich2017dna}, the reliability of each received symbol is proportional to its number of occurrence, which can be utilized by the BFA.

\section{Implementation of Basis-Finding Algorithm}\label{section: implementation}

While Algorithm \ref{algo: BFA} provides a general outline for the BFA, we develop three ways to  implement the BFA in detail in this section.
The first way is called straightforward implementation, as it is based on the conventional Gaussian elimination.
The other two ways employ a pre-processing step to rearrange the received symbols before applying the straightforward implementation.
More specifically, motivated by Observation 1, the second way,  referred to as the sorted-weight implementation, first sorts the received symbols in descending order according to their weights and then applies the straightforward implementation.
The third way, named as the  triangulation-based implementation, employs triangulation (the key idea of inactivation decoding  \cite{odlyzko1984discrete, lamacchia1991solving, he2020disjoint, lazaro2017fountain,  3GPP06, shokrollahi2005systems, shokrollahi2011raptor}) to reduce the complexity of the straightforward implementation. It also involves a weight optimization in the triangulation process due to Observation 1.

\subsection{Straightforward Implementation}\label{subsection: straightforward implementation}

The idea is to use Gaussian elimination to realize (W1) and (W2) so as to generate $B(\mb{A}, \mb{Y})$ and to identify the LRs each basis element of $B(\mb{A}, \mb{Y})$ attends.
Let $r$ denote the number of existing basis elements, which equals to zero at the beginning and increases by one when a new basis element is added into $B(\mb{A}, \mb{Y})$.
For any $i \in [m]$, denote $(\mb{A}{[i]}, \mb{Y}{[i]})$ as the first $i$ received symbols.
We always maintain $(\mb{A}{[r]}, \mb{Y}{[r]})$ as the basis elements that are found (i.e., maintain $B(\mb{A}, \mb{Y}) = (\mb{A}{[r]}, \mb{Y}{[r]})$) through applying necessary row-exchange to $(\mb{A}, \mb{Y})$.
Moreover, we maintain an upper triangular form of $(\mb{A}{[r]}, \mb{Y}{[r]})$, denoted by $\mb{B}$, and use a matrix $\mb{Q}$ to record the elementary row operations leading $(\mb{A}{[r]}, \mb{Y}{[r]})$ to $\mb{B}$.
Specifically, for the $i$-th row of $\mb{B}$ (resp. $\mb{Q}$), denoted by $\mb{b}_i$ (resp. $\mb{q}_i$), we always maintain $\mb{b}_i = \mb{q}_i \cdot (\mb{A}{[r]}, \mb{Y}{[r]})$ if both $\mb{b}_i \neq \mr{NULL}$ and $\mb{q}_i \neq \mr{NULL}$ (i.e., if $\mb{b}_i$ and $\mb{q}_i$ are valid).
The matrix $\mb{B}$ (instead of $(\mb{A}{[r]}, \mb{Y}{[r]})$) is used to eliminate the non-zero entries in a received symbol, while $\mb{Q}$ is used to track the LR of the received symbol.
For any vector $\mb{b}$ over $\mb{F}_2$, we use $\psi(\mb{b})$ to denote the index of the leftmost non-zero entry of $\mb{b}$, and let $\psi(\mb{b}) = \infty$ if $\mb{b}$ is a zero vector.
We summarize the corresponding decoding process in Algorithm \ref{algo: straightforward implementation}.

\begin{algorithm}[t!]
    \caption{Straightforward implementation of basis-finding algorithm}
    \label{algo: straightforward implementation}
    \begin{algorithmic}[1]
        \REQUIRE    $(\mb{A}, \mb{Y})$.
        \ENSURE     $\hat{\mb{X}}$.
        \STATE  $r \leftarrow 0$.
        \STATE  $\mb{b}_i \leftarrow \mb{q}_i \leftarrow \mr{NULL}, \forall i \in [n+l]$. $//$There are at most $n+l$ basis elements.

        \FOR {$i \leftarrow 1, 2, \ldots, m$}\label{code: straightforward implementation @ for i}
            \STATE  $\mb{b} \leftarrow [b_j]_{1 \leq j \leq n+l} \leftarrow (\mb{a}_i, \mb{y}_i)$.
            \STATE  $\mb{q} \leftarrow [q_j]_{1 \leq j \leq r} \leftarrow 0^r$. $//$Set $\mb{b} =  \mb{q} \cdot (\mb{A}{[r]}, \mb{Y}{[r]}) \oplus (\mb{a}_i, \mb{y}_i)$.
            \WHILE  {$\psi(\mb{b}) \neq \infty$}\label{code: straightforward implementation @ while}
                \STATE  $t \leftarrow \psi(\mb{b})$.
                \IF {$\mb{b}_{t} \neq \mr{NULL}$}
                    \STATE  $\mb{b} \leftarrow \mb{b} \oplus \mb{b}_{t}$. $//$Make $b_{t} = 0$ and then $\psi(\mb{b})$ must increase.\label{code: straightforward implementation @ update b}
                    \STATE  $\mb{q} \leftarrow \mb{q} \oplus \mb{q}_{t}$. $//$Maintain $\mb{b} = \mb{q} \cdot (\mb{A}{[r]}, \mb{Y}{[r]})\oplus (\mb{a}_i, \mb{y}_i)$ since $\mb{b}_{t} = \mb{q}_{t} \cdot (\mb{A}{[r]}, \mb{Y}{[r]})$.\label{code: straightforward implementation @ update q}
                \ELSE
                    \STATE  $//$$(\mb{a}_i, \mb{y}_i)$ is a basis element.
                    \STATE  $r \leftarrow r + 1$.
                    \STATE  Swap $(\mb{a}_i, \mb{y}_i)$ and $(\mb{a}_r, \mb{y}_r)$. $//$Maintain $(\mb{A}{[r]}, \mb{Y}{[r]})$ as the basis of $(\mb{A}{[i]}, \mb{Y}{[i]})$.
                    \STATE  $N_r = 0$. $//$Number of LRs $(\mb{a}_r, \mb{y}_r)$ attends.
                    \STATE  $\mb{b}_{t} \leftarrow \mb{b}$.
                    \STATE  $\mb{q}_{t} \leftarrow [\mb{q} ~ 1]$. $//\mb{b}_{t} = \mb{q}_{t} \cdot (\mb{A}{[r]}, \mb{Y}{[r]})$ is kept.
                    \STATE  For any $t \neq t' \in [n+l]$ with $\mb{q}_{t'} \neq \mr{NULL}$, append one zero to the tail of $\mb{q}_{t'}$. $//$Maintain $\mb{b}_{t'} = \mb{q}_{t'} \cdot (\mb{A}{[r]}, \mb{Y}{[r]})$.
                    \STATE  Goto \ti{label\_out}. $//$Goto line \ref{code: straightforward implementation @ label out}.
                \ENDIF
            \ENDWHILE

            \STATE  $N_{j} \leftarrow N_j + 1, \forall j \in [r]$ with $q_j = 1$. $//\mb{b}$ is a zero vector, indicating $(\mb{a}_i, \mb{y}_i) = \oplus_{j \in [r], q_j = 1} (\mb{a}_j, \mb{y}_j)$.
            \STATE  \ti{label\_out}.\label{code: straightforward implementation @ label out}
        \ENDFOR
        \STATE  $//$At this point, we have $B(\mb{A}, \mb{Y}) = (\mb{A}{[r]}, \mb{Y}{[r]})$,  and for $i \in [r]$, $(\mb{a}_i, \mb{y}_i)$ attends $N_i$ many LRs of $(\mb{A}, \mb{Y}) \setminus B(\mb{A}, \mb{Y})$.
        \IF {$\exists N^* \geq 0$ with $|\{i \in [r]: N_i \geq N^*\}| = n$}\label{code: straightforward implementation @ N*}
            \STATE  Let $B_n(\mb{A}, \mb{Y})$ consist of the rows of $(\mb{A}, \mb{Y})$ with indices $\{i \in [r]: N_i \geq N^*\}$.
            \STATE  Recover $\mb{X}$: Give an estimation $\hat{\mb{X}}$ of $\mb{X}$ by applying the inactivation decoding \cite{odlyzko1984discrete, lamacchia1991solving, he2020disjoint, lazaro2017fountain,  3GPP06, shokrollahi2005systems, shokrollahi2011raptor} on $B_n(\mb{A}, \mb{Y})$.
        \ELSE
            \STATE  Claim a decoding failure.
        \ENDIF
    \end{algorithmic}
\end{algorithm}

\begin{example}

Take \eqref{eqn: eg A, y} as an example,  and assume that only $(\mb{a}_1, \mb{y}_1)$ and $(\mb{a}_5, \mb{y}_5)$ are incorrect.
At the end of Algorithm \ref{algo: straightforward implementation}, $(\mb{A}, \mb{Y})$ does not change, $(\mb{A}[3], \mb{Y}[3])$ forms the basis ($r = 3$), and $\mb{B}$ and $\mb{Q}$ become
\begin{equation*}
    [\mb{B} \mid \mb{Q}] =
    \left[
        \begin{array}{c c c c | c c c}
            1 & 1 & 0 & 1 & 1 & 0 & 0\\
            0 & 1 & 1 & 0 & 1 & 1 & 0\\
            0 & 0 & 1 & 1 & 1 & 0 & 1\\
            - & - & - & - & - & - & -\\
        \end{array}
    \right],
\end{equation*}
where ``$-$" indicates the fourth row of $\mb{B}$ and $\mb{Q}$ equal to $\mr{NULL}$.
$\mb{B}$ is an upper triangular matrix which can be used to eliminate the non-zero entries of a received symbol in a straightforward way, while $\mb{Q}$ can be used to identify the LC of $(\mb{A}[r], \mb{Y}[r])$ that represents the received symbol.
More specifically,  we can easily see that $(\mb{a}_4, \mb{y}_4) = \mb{b}_2 \oplus \mb{b}_3$.
Accordingly, we have $(\mb{a}_4, \mb{y}_4) = (\mb{q}_2 \oplus \mb{q}_3) \cdot (\mb{A}[3], \mb{Y}[3]) = (0, 1, 1) \cdot (\mb{A}[3], \mb{Y}[3]) = (\mb{a}_2, \mb{y}_2) \oplus (\mb{a}_3, \mb{y}_3)$.
On the other hand, we can similarly have $(\mb{a}_5, \mb{y}_5) = \mb{b}_1 \oplus \mb{b}_2 \oplus \mb{b}_3 = (\mb{a}_1, \mb{y}_1)  \oplus (\mb{a}_2, \mb{y}_2) \oplus (\mb{a}_3, \mb{y}_3)$.
As a result, $B_n(\mb{A}, \mb{Y})$ consists of  $(\mb{a}_2, \mb{y}_2)$ and $(\mb{a}_3, \mb{y}_3)$, which are then used to successfully recover $\mb{X}$.
\end{example}

We remark that the condition in line \ref{code: straightforward implementation @ N*} of Algorithm \ref{algo: straightforward implementation} is to ensure a unique way to form $B_n(\mb{A}, \mb{Y})$.
Otherwise, there will be no way to form $B_n(\mb{A}, \mb{Y})$ if $r < n$, or there will be $\binom{u - v}{n - v}$ ways if $\exists N^* \geq 0$ with $u = |\{i \in [r]: N_i \geq N^*\}| > n$ and $v = |\{i \in [r]: N_i \geq N^* + 1\}| < n$.
Assuming there are $\binom{u - v}{n - v}$ ways to form $B_n(\mb{A}, \mb{Y})$, the probability for $B_n(\mb{A}, \mb{Y})$ to consist of $n$ correct basis elements is at most $1 / \binom{u - v}{n - v} \leq 1/2$ if $(\mb{A}{[r]}, \mb{Y}{[r]})$ contains $n$ correct basis elements and is $0$ otherwise.
This indicates that if the condition in line \ref{code: straightforward implementation @ N*} of Algorithm \ref{algo: straightforward implementation} does not hold, we are not able to recover $\mb{X}$ with a reasonable reliability.
Therefore, we claim a decoding failure for this situation.

The most time consuming part of Algorithm \ref{algo: straightforward implementation} lies on lines \ref{code: straightforward implementation @ update b} and \ref{code: straightforward implementation @ update q}.
For a given $i$ of line \ref{code: straightforward implementation @ for i}, lines \ref{code: straightforward implementation @ update b} and \ref{code: straightforward implementation @ update q} can be implemented for at most $r$ times, each time with complexity $O(n + l)$, where $r = \mr{rank}(\mb{A}, \mb{Y})$ is the upper bound of the number of valid $\mb{b}_t$ for line \ref{code: straightforward implementation @ update b}.
Thus, the time complexity of Algorithm \ref{algo: straightforward implementation} is given by $O(m r (n+l))$, or $O(m (n+l)^2)$ for simplicity.
We remark that the time complexity for recovering $\mb{X}$ from $B_n(\mb{A}, \mb{Y})$ is $O(n^2(n+l))$ if the conventional Gaussian elimination is used, and may be reduced if the inactivation decoding \cite{odlyzko1984discrete, lamacchia1991solving, he2020disjoint, lazaro2017fountain,  3GPP06, shokrollahi2005systems, shokrollahi2011raptor} is applied.
Thus, this time complexity does not dominate that of Algorithm \ref{algo: straightforward implementation}.

\subsection{Sorted-Weight Implementation}

Observation 1 motivates us to propose a sorted-weight implementation.
The idea is to first sort the received symbols in descending order according to their weights and then apply the straightforward implementation.
Doing so is expected to generate a basis with larger average weight of its basis elements than directly applying the straightforward implementation.
Obviously, the sorted-weight implementation has the same order of complexity  as the straightforward implementation, i.e., $O(r m (n+l))$, since the complexity caused by sorting ($O(m \log m)$) does not dominate the overall decoding complexity.

\begin{example}
We take the received symbols in \eqref{eqn: eg A, y} as the toy example to show how the sorted-weight implementation works.
Assume that only $(\mb{a}_1, \mb{y}_1)$ and $(\mb{a}_5, \mb{y}_5)$ are incorrect.
First, we sort the received symbols in descending order according to their weights (number of ones in each row of $\mb{A} \in \mbb{F}_2^{5 \times 2}$), leading to
\begin{equation*}
    (\mb{A}, \mb{Y}) =
    \left[
        \begin{array}{c c c c}
            1 & 1 & 0 & 1\\
            1 & 1 & 1 & 0\\
            1 & 0 & 1 & 1\\
            0 & 1 & 0 & 1\\
            1 & 0 & 0 & 0\\
        \end{array}
    \right],
\end{equation*}
where only the originally second and third received symbols are swapped.
Then, after calling Algorithm \ref{algo: straightforward implementation}, $(\mb{A}, \mb{Y})$ does not change, and $\mb{B}$ and $\mb{Q}$ become
\begin{equation*}
    [\mb{B} \mid \mb{Q}] =
    \left[
        \begin{array}{c c c c | c c c}
            1 & 1 & 0 & 1 & 1 & 0 & 0\\
            0 & 1 & 1 & 0 & 1 & 0 & 1\\
            0 & 0 & 1 & 1 & 1 & 1 & 0\\
            - & - & - & - & - & - & -\\
        \end{array}
    \right].
\end{equation*}
By making use of $\mb{B}$ and $\mb{Q}$, we can easily see that $(\mb{a}_4, \mb{y}_4) = (\mb{a}_2, \mb{y}_2) \oplus (\mb{a}_3, \mb{y}_3)$ and $(\mb{a}_5, \mb{y}_5) = (\mb{a}_1, \mb{y}_1)  \oplus (\mb{a}_2, \mb{y}_2) \oplus (\mb{a}_3, \mb{y}_3)$.
As a result, $B_n(\mb{A}, \mb{Y})$ consists of $(\mb{a}_2, \mb{y}_2)$ and $(\mb{a}_3, \mb{y}_3)$, which are then used to correctly recover $\mb{X}$.
\end{example}

\subsection{Triangulation-Based Implementation}

Let $\theta$ denote the average weight of each row of $\mb{A}$.
For practical fountain codes, $\mb{A}$ is very sparse, i.e., $\theta$ is very small compared to $n$.
For example, LT codes \cite{luby2002lt} have $\theta = O(\log n)$ and Raptor codes \cite{	shokrollahi2006raptor} have $\theta = O(1)$.
As a result, we can efficiently implement Algorithm \ref{algo: BFA} by making use of the sparsity of $\mb{A}$.
The key idea is to first triangulate $\mb{A}$, similar to the inactivation decoding \cite{odlyzko1984discrete, lamacchia1991solving, he2020disjoint, lazaro2017fountain,  3GPP06, shokrollahi2005systems, shokrollahi2011raptor}.
In this section, we illustrate this triangulation-based implementation for Algorithm \ref{algo: BFA}.

The triangulation-based implementation contains two steps.
First, convert $(\mb{A}, \mb{Y})$ to the triangular form given below:
\begin{equation}\label{eqn: L R}
(\mb{A}, \mb{Y}) = [\mb{L} ~~ \mb{R}]
\end{equation}
through necessary row-exchange and column-exchange, where $\mb{L} = [l_{i, j}]_{1 \leq i \leq m, 1 \leq j \leq \gamma} \in \mbb{F}_2^{m \times \gamma}$ is a lower triangular matrix, i.e., $l_{i, j} = 0$ for $1 \leq i < j \leq \gamma$ and $l_{i, i} \neq 0$ for $1 \leq i \leq \gamma$,
and $\mb{R} \in \mbb{F}_2^{m \times (n+l-\gamma)}$ consists of the remaining columns of $(\mb{A}, \mb{Y})$.
Second, call Algorithm \ref{algo: straightforward implementation} with \eqref{eqn: L R} as input.

It has been well studied in \cite{odlyzko1984discrete, lamacchia1991solving, he2020disjoint, lazaro2017fountain,  3GPP06, shokrollahi2005systems, shokrollahi2011raptor} on how to efficiently get the form of \eqref{eqn: L R}.
The most advanced implementation can be found in \cite{ he2020disjoint}.
We would like to highlight that according to Observation 1, we include a weight optimization operation in the triangulation process by selecting received symbols having larger weights with higher priority to form the top rows  of \eqref{eqn: L R} if multiple candidates are available.
The columns contained in $\mb{R}$ are called the inactive columns.
In particular, motivated by the idea of permanent inactivation \cite{shokrollahi2011raptor}, we always let $\mb{R}$ contain $\mb{Y}$ and possibly a few columns of $\mb{A}$ ($\gamma = n - o(n)$),  since $\mb{Y}$ is uniformly chosen from $\mbb{F}_2^{m \times l}$ which thus is generally very dense and can hardly attend $\mb{L}$.
We can start to analyze the complexity of Algorithm \ref{algo: straightforward implementation} given the form of \eqref{eqn: L R} from the following lemma.

\begin{lemma}\label{lemma: efficiency}
The first $\gamma$ entries of the $\mb{b}_t$ in line \ref{code: straightforward implementation @ update b} of Algorithm \ref{algo: straightforward implementation} are zeros except that the $t$-th entry is non-zero for $t \leq \gamma$ (i.e. after handling $(\mb{A}[\gamma], \mb{Y}[\gamma])$, the first $\gamma$ rows and columns of $\mb{B}$ form an identity matrix).
\end{lemma}

\begin{IEEEproof}
It is easy to verify Lemma \ref{lemma: efficiency} by noting that $\mb{L}$ is a lower triangular matrix and the rows of $[\mb{L} ~ \mb{R}]$ are handled from top to bottom in order.
\end{IEEEproof}

For example, we convert \eqref{eqn: eg A, y} to the form of \eqref{eqn: L R}, say
\begin{equation}\label{eqn: eg A, y form L R}
    (\mb{A}, \mb{Y}) = [\mb{L} ~~ \mb{R}] =
    \left[
        \begin{array}{c c c c}
            1 & 0 & 1 & 1\\
            1 & 1 & 0 & 1\\
            1 & 1 & 1 & 0\\
            0 & 1 & 0 & 1\\
            1 & 0 & 0 & 0\\
        \end{array}
    \right],
\end{equation}
where $\gamma = n = 2$.
Then, after calling Algorithm \ref{algo: straightforward implementation}, $(\mb{A}, \mb{Y})$ does not change, and $\mb{B}$ and $\mb{Q}$ become
\begin{equation}\label{eqn: eg A, y form L R, B Q}
    [\mb{B} \mid \mb{Q}] =
    \left[
        \begin{array}{c c c c | c c c}
            1 & 0 & 1 & 1 & 1 & 0 & 0\\
            0 & 1 & 1 & 0 & 1 & 1 & 0\\
            0 & 0 & 1 & 1 & 0 & 1 & 1\\
            - & - & - & - & - & - & -\\
        \end{array}
    \right].
\end{equation}
We can see that Lemma \ref{lemma: efficiency} is satisfied.

Given the form of \eqref{eqn: L R}, we are able to reduce the complexity of the most time consuming part of Algorithm \ref{algo: straightforward implementation}, i.e., lines \ref{code: straightforward implementation @ update b} and \ref{code: straightforward implementation @ update q}.
Specifically, for a given $i$ of line \ref{code: straightforward implementation @ for i}, lines \ref{code: straightforward implementation @ update b} and \ref{code: straightforward implementation @ update q} are implemented by at most $\min(r, \theta_i + n - \gamma + l)$ times, where $r = \mr{rank}(\mb{A}, \mb{Y})$ and $\theta_i$ is the number of non-zero entries among the first $\gamma$ entries of $(\mb{a}_i, \mb{y}_i)$.
The term $\theta_i + n - \gamma + l$ is the upper bound of the times for running the while loop of line \ref{code: straightforward implementation @ while},  since no new non-zero entries occur to the first $\gamma$ entries of $\mb{b}$ according to Lemma \ref{lemma: efficiency}.
On the other hand, at each time lines \ref{code: straightforward implementation @ update b} and \ref{code: straightforward implementation @ update q} have time complexity $O(n+l)$.
Therefore, given the form of \eqref{eqn: L R}, the time complexity of Algorithm \ref{algo: straightforward implementation} becomes $O(\min(r, \theta + n - \gamma + l)m(n + l))$.

In practice, we have a high chance to make $\gamma = n - O(1)$ for reasonable values of $(p, n, l, m)$.
Moreover, the time complexity for getting the form of \eqref{eqn: L R} is negligible compared to that of Algorithm \ref{algo: straightforward implementation}.
Thus, the time complexity of the proposed triangulation-based implementation is $O(\min(r, \theta + n - \gamma + l)m(n + l))$, which can be significantly smaller than $O(r m (n+l))$, the time complexity of the straightforward implementation (not given the form of \eqref{eqn: L R}), particularly for $r \geq n \gg l$.

\begin{example}
\label{example: triangulation}
We take \eqref{eqn: eg A, y} as the toy example to show how the triangulation-based implementation works.
Assume that only $(\mb{a}_1, \mb{y}_1)$ and $(\mb{a}_5, \mb{y}_5)$ are incorrect.
First, we convert \eqref{eqn: eg A, y} to the triangular form of \eqref{eqn: eg A, y form L R}, where only the first two rows of \eqref{eqn: eg A, y} are swapped (accordingly, $(\mb{a}_1, \mb{y}_1)$ becomes correct and $(\mb{a}_2, \mb{y}_2)$ becomes incorrect).
Then, after calling Algorithm \ref{algo: straightforward implementation}, $(\mb{A}, \mb{Y})$ does not vary from \eqref{eqn: eg A, y form L R}, $(\mb{A}[3], \mb{Y}[3])$ forms the basis,  and $\mb{B}$ and $\mb{Q}$ are given by \eqref{eqn: eg A, y form L R, B Q}.
By making use of $\mb{B}$ and $\mb{Q}$, we can easily see that $(\mb{a}_4, \mb{y}_4) = (\mb{a}_1, \mb{y}_1) \oplus (\mb{a}_3, \mb{y}_3)$ and $(\mb{a}_5, \mb{y}_5) = (\mb{a}_1, \mb{y}_1)  \oplus (\mb{a}_2, \mb{y}_2) \oplus (\mb{a}_3, \mb{y}_3)$.
As a result, $B_n(\mb{A}, \mb{Y})$ consists of $(\mb{a}_1, \mb{y}_1)$ and $(\mb{a}_3, \mb{y}_3)$, which are then used to correctly recover $\mb{X}$.
\end{example}

We would like to highlight that the weight optimization based on Observation 1 is also important for the triangulation-based implementation, as it enables  a larger average weight of the basis elements and hence the better error-correction performance.
We use the following toy example to  illustrate this point.

\begin{example}
Similar to Example \ref{example: triangulation}, we consider the received symbols given by \eqref{eqn: eg A, y} and assume that only $(\mb{a}_1, \mb{y}_1)$ and $(\mb{a}_5, \mb{y}_5)$ are incorrect.
Instead of converting \eqref{eqn: eg A, y} to the triangular form of \eqref{eqn: eg A, y form L R}, we can also convert \eqref{eqn: eg A, y} to the following triangular form without the weight optimization:
\begin{equation}
\label{eqn: triangulation without optimization}
    (\mb{A}, \mb{Y}) =
    \left[
        \begin{array}{c c c c}
            1 & 0 & 1 & 1\\
            0 & 1 & 0 & 1\\
            1 & 1 & 0 & 1\\
            1 & 1 & 1 & 0\\
            1 & 0 & 0 & 0\\
        \end{array}
    \right].
\end{equation}
The second received symbol in \eqref{eqn: triangulation without optimization} has smaller weight than that in \eqref{eqn: eg A, y form L R} (i.e., 1 vs 2).
Then, after calling Algorithm \ref{algo: straightforward implementation}, $(\mb{A}, \mb{Y})$ does not change, and $\mb{B}$ and $\mb{Q}$ become
\begin{equation*}
    [\mb{B} \mid \mb{Q}] =
    \left[
        \begin{array}{c c c c | c c c}
            1 & 0 & 1 & 1 & 1 & 0 & 0\\
            0 & 1 & 0 & 1 & 0 & 1 & 0\\
            0 & 0 & 1 & 1 & 1 & 1 & 1\\
            - & - & - & - & - & - & -\\
        \end{array}
    \right].
\end{equation*}
By making use of $\mb{B}$ and $\mb{Q}$, we can easily see that $(\mb{a}_4, \mb{y}_4) = (\mb{a}_1, \mb{y}_1) \oplus (\mb{a}_2, \mb{y}_2)$ and $(\mb{a}_5, \mb{y}_5) = (\mb{a}_2, \mb{y}_2) \oplus (\mb{a}_3, \mb{y}_3)$.
In this case, the condition in line \ref{code: straightforward implementation @ N*} of Algorithm \ref{algo: straightforward implementation} is not satisfied, i.e., there are multiple ways to form $B_n(\mb{A}, \mb{Y})$.
As a result, Algorithm \ref{algo: straightforward implementation} claims a decoding failure.
\end{example}

\subsection{Remarks}

Recall that the straightforward implementation generates a basis in the way of (W1) and (W2), where the received symbols $(\mb{a}_1, \mb{y}_1)$, $(\mb{a}_2, \mb{y}_2)$, $\ldots$, $(\mb{a}_m, \mb{y}_m)$ are handled one-by-one in order.
Obviously, the average weight of the basis elements for the straightforward implementation should be very close to the weight expectation of a received symbol.
On the other hand, by reordering the received symbols from the largest weight to the smallest weight, the sorted-weight implementation can have larger average weight, since received symbols with larger weights have higher priority to be selected and added into the basis.
For the triangulation-based implementation, the received symbols are rearranged to form the triangular form of \eqref{eqn: L R}.
By selecting received symbols having larger weights with higher priority to form the top rows  of \eqref{eqn: L R},  the triangulation-based implementation can have larger average weight than that of the straightforward implementation.
Moreover, among the three implementations, the triangulation-based implementation has the lowest complexity.

\section{Error-Correction Performance Analysis}\label{section: performance analysis}

In this section, we analyze the error-correction performance of the proposed BFA given by Algorithm \ref{algo: BFA}.
For the ease of analysis, we use the following three assumptions throughout this section.
However, these assumptions are not necessary for the BFA to apply.
\begin{enumerate}[(i)]
\item   Random fountain codes are considered, i.e., each entry of $\mb{A}$ is independently and uniformly chosen from $\mbb{F}_2$.

\item   Each incorrect received symbol suffers from uniformly random substitution errors, i.e., the assumption of \eqref{eqn: random error}.
\item   The straightforward implementation proposed in Section \ref{subsection: straightforward implementation}, i.e., Algorithm \ref{algo: straightforward implementation}, is employed.\footnote{ In this case, the received symbols $(\mb{a}_1, \mb{y}_1)$, $(\mb{a}_2, \mb{y}_2)$, $\ldots$, $(\mb{a}_m, \mb{y}_m)$ are handled one-by-one in order, and each received symbols are independent and identically distributed (i.i.d.). On the contrary, if the received symbols are rearranged according to the sorted-weight implementation or triangulation-based implementation, consecutive received symbols are not i.i.d.. As a result, it would be hard to quantitatively analyze the behavior of a basis (e.g., compute the probability that a basis contains $n$ correct received symbols).}
\end{enumerate}

For $i \in [m]$, recall that $(\mb{a}_i, \mb{y}_i)$ denotes the $i$-th received symbol and $(\mb{A}[i], \mb{Y}[i])$ denotes the first $i$ received symbols.
For convenience, we use $B(i)$ to refer to $B(\mb{A}[i], \mb{Y}[i])$, the basis that is formed by Algorithm \ref{algo: straightforward implementation} after handling $(\mb{A}[i], \mb{Y}[i])$.
In particular, we let $B(0) = \emptyset$.
Denote $B_c(i)$ and $B_e(i)$ as the sets of correct and incorrect basis elements of $B(i)$, respectively.
We have $B(i) = B_c(i) \cup B_e(i)$ and $B_c(i) \cap B_e(i) = \emptyset$.
Moreover, we define the following events.
\begin{itemize}
    \item   $E$: the event that $|B_c(m)| = n$, i.e., $B(\mb{A}, \mb{Y})$ has  $n$  correct basis elements.
        ($E$ is necessary for the BFA to successfully recover $\mb{X}$.)
    \item   $E_1$: the event that $(\mb{A}, \mb{Y})$ contains $n$ linearly independent correct received symbols.
        ($E_1$ is necessary for $E$ to occur.)
    \item   $E_2$: the event that the error patterns of $B_e(m)$ are linearly independent.
        ($E_2$ is necessary for $E$ to occur.)
    \item   $E_3$: the event that the error patterns of the incorrect received symbols of $(\mb{A}, \mb{Y})$ are linearly independent.
        ($E_3$ is sufficient for $E_2$ to occur.)
    \item   $F$: the event that each basis element of $B_c(m)$ attends more LRs of $(\mb{A}, \mb{Y}) \setminus B(\mb{A}, \mb{Y})$ than each basis element of $B_e(m)$.
        ($F$ is necessary for the BFA to successfully recover $\mb{X}$.)
    \item   $G$: the joint event of $E_1, E_3$, and $F$, i.e., $(E_1, E_3, F)$.
        ($G$ is sufficient for the BFA to successfully recover $\mb{X}$.)
\end{itemize}

\begin{lemma}\label{lemma: (E, F)}
The event $(E, F)$ is equivalent to that Algorithm \ref{algo: straightforward implementation} correctly recovers $\mb{X}$.
\end{lemma}

\begin{IEEEproof}
If $(E, F)$ happens, it is easy to see that $B_n(\mb{A}, \mb{Y})$ generated by Algorithm \ref{algo: straightforward implementation} must be equal to $B_c(m)$, which consists of $n$ correct basis elements, indicating that Algorithm \ref{algo: straightforward implementation} can correctly recover $\mb{X}$.
On the other hand, if Algorithm \ref{algo: straightforward implementation} correctly recovers $\mb{X}$, $B_n(\mb{A}, \mb{Y})$ generated by Algorithm \ref{algo: straightforward implementation} must consist of $n$ correct basis elements.
As a result, $(E, F)$ must happen.
This completes the proof.
\end{IEEEproof}

With respect to Lemma \ref{lemma: (E, F)}, our task is to investigate $\mbb{P}(E, F)$.
However, it is hard to find a practical method to explicitly compute $\mbb{P}(E, F)$.
Instead, in the following, we first analyze $\mbb{P}(E)$, which is an upper bound for $\mbb{P}(E, F)$ such that
\[
    \mbb{P}(E) \geq \mbb{P}(E, F).
\]
Then, we analyze $\mbb{P}(G)$, which is expected to be a tight lower bound for $\mbb{P}(E, F)$ for relatively small to moderate values of $m$ (explained later).
Finally, we analyze $\mbb{P}(E, F)$ for relatively large $m$.
To start, we present Lemma \ref{lemma: hoeffding inequality} below which will be extensively used in the subsequent analysis.

\begin{lemma}\label{lemma: hoeffding inequality}
Consider a coin that shows heads with probability $p$ and tails with probability $1 - p$.
We toss the coin $t$ times and denote $U(t)$ as the number of times the coin comes up heads.
We have
\[
    \mbb{P}(U(t) = i) = \binom{t}{i}p^i(1-p)^{t-i}, \quad i = 0, 1, \ldots, t.
\]
Moreover, for any given $\epsilon > 0$, we have
\begin{align*}
    &\mbb{P}(U(t) \leq (p - \epsilon)t) \leq e^{-2\epsilon^2t} \text{~and}\\
    &\mbb{P}(U(t) \geq (p + \epsilon)t) \leq e^{-2\epsilon^2t}.
\end{align*}
\end{lemma}

\begin{IEEEproof}
The lemma is obviously correct once we note that $U(t)$ follows the binomial distribution and the bounds are  direct results of the Hoeffding's inequality \cite{hoeffding1963probability}.
\end{IEEEproof}
%
%

\subsection{Analysis of $\mbb{P}(E)$}\label{subsection: P(E)}

\begin{lemma}\label{lemma: E E1 E2}
($E_1, E_2$) is equivalent to $E$.
\end{lemma}

\begin{IEEEproof}
Assume that $E$ happens.
Then, $E_1$ must happen.
Moreover, if $E_2$ does not happen, the basis elements of $B(m)$ are not linearly independent, leading to a contradiction.
Thus, both $E_1$ and $E_2$ must happen.
On the other hand, assume that both $E_1$ and $E_2$ happen. It is obvious that $E$ happens.
\end{IEEEproof}

Lemma \ref{lemma: E E1 E2} leads to two necessary conditions for $(p, n, l, m)$ in order to ensure a sufficiently high $\mbb{P}(E)$.
On the one hand, it is necessary to have
\begin{equation}\label{eqn: m > n/p}
    m > n / p,
\end{equation}
otherwise $\mbb{P}(E_1)$ cannot be sufficiently high according to Lemmas \ref{lemma: hoeffding inequality} and \ref{lemma: rank property}.
On the other hand, it is also necessary to have
\begin{equation}\label{eqn: l > n(1-p)/p}
    l > \frac{1 - p}{p} n.
\end{equation}
Otherwise, we have $n + l \leq n / p < m$.
Then, $(\mb{A}[n+l], \mb{Y}[n+l])$ will have a high chance to include less than $n$ correct and more than $l$ incorrect received symbols according to Lemma \ref{lemma: hoeffding inequality}.
Moreover, according to Lemma \ref{lemma: rank property}, these incorrect received symbols are very likely to contain more than $l$ linearly independent received symbols.
As a result, $|B_e(n+l)| > l$ happens with a high chance, indicating that $\mbb{P}(E_2)$ as well as $\mbb{P}(E)$ cannot be sufficiently high.

Given $(p, n, l, m)$, we can explicitly compute $\mbb{P}(E)$ by using dynamic programming \cite[Section 15.3]{introAlgo01}.
To this end, for $0 \leq i \leq m, 0 \leq n_c \leq n,$ and $0 \leq n_e \leq l$, define $P_E(i, n_c, n_e)$ as the probability that $|B_c(i)| = n_c, |B_e(i)| = n_e$, and the error patterns of $B_e(i)$ are linearly independent.
We require the error patterns of $B_e(i)$ to be linearly independent so as to ensure $E_2$.
We have the following theorem.

\begin{theorem}\label{theorem: P(E) = sum P_E}
We have
\begin{equation}\label{eqn: P(E) = sum P_E}
    \mbb{P}(E) = \sum_{0 \leq n_e \leq l} P_E(m, n, n_e).
\end{equation}
Moreover, we have $P_E(0, 0, 0) = 1$, and for $0 <  i \leq m, 0 \leq n_c \leq n,$ and $0 \leq n_e \leq l$, we have
\begin{align}\label{eqn: P_E}
    P_E&(i, n_c, n_e) = P_E(i - 1, n_c, n_e) p \frac{2^{n_c}}{2^n} +\nonumber
    \\& P_E(i - 1, n_c, n_e) (1 - p) \frac{2^{n_c + n_e} - 2^{n_c}}{2^{n+l} - 2^n} +\nonumber
    \\&P_E(i - 1, n_c - 1, n_e) p \frac{2^n - 2^{n_c - 1}}{2^n} +\nonumber
    \\& P_E(i - 1, n_c, n_e - 1) (1 - p) \frac{2^{n + l} - 2^{n + n_e - 1}}{2^{n+l} - 2^n},
\end{align}
where for simplicity, we let $P_E(\cdot, n_c', n_e') = 0$ for any $n_c' < 0$ or $n_e' < 0$.
\end{theorem}

\begin{IEEEproof}
See Appendix \ref{appendix: P(E) = sum P_E}.
\end{IEEEproof}

\begin{figure}[t]
\centering
\includegraphics[scale = 0.5]{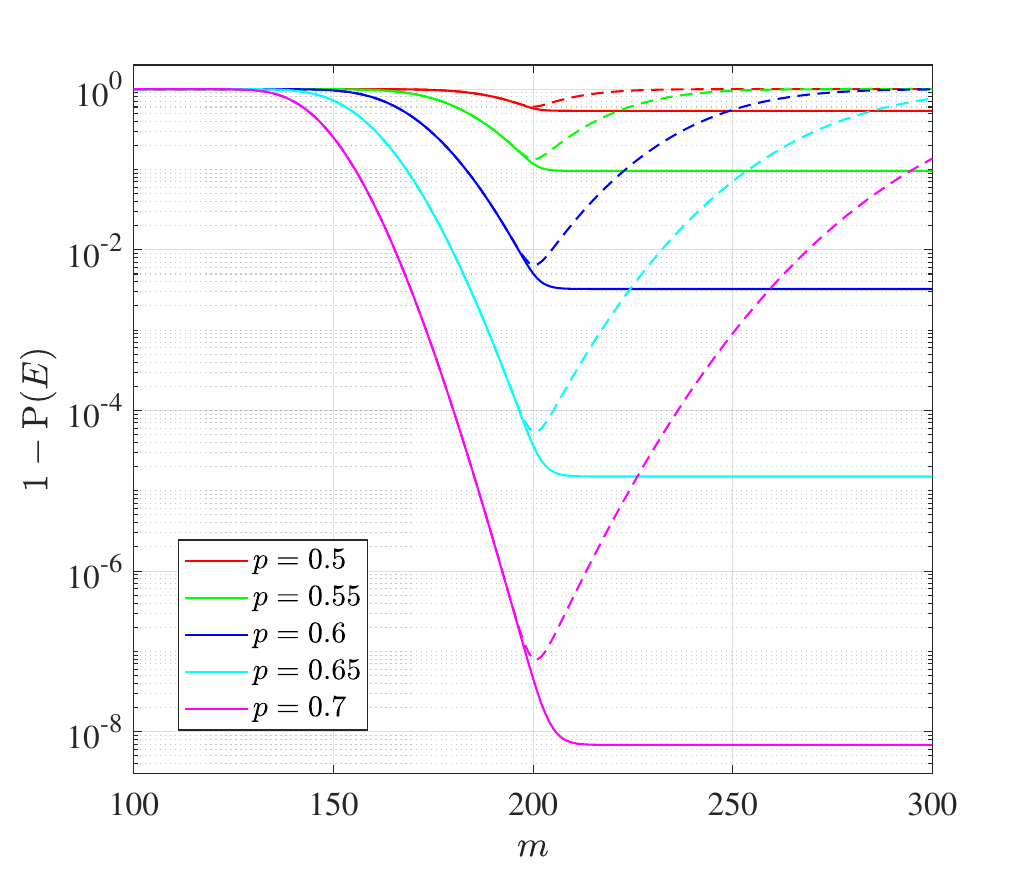}
\caption{Numerical result of $1 - \mbb{P}(E)$ for $(n, l) = (100, 100)$, where solid and dash lines are computed based on \eqref{eqn: P(E) = sum P_E} and \eqref{eqn: P(E) > P(E1, E3)}, respectively.}
\label{fig: PE}
\end{figure}

According to \eqref{eqn: P_E}, we can compute $P_E(i, n_c, n_e)$ for all $0 <  i \leq m, 0 \leq n_c \leq n,$ and $0 \leq n_e \leq l$ with time complexity $O(mnl)$.
Then, we can compute $\mbb{P}(E)$ with time complexity $O(l)$ based on \eqref{eqn: P(E) = sum P_E}.
As an example, we show the numerical result of $1 - \mbb{P}(E)$ in Fig. \ref{fig: PE} for $(n, l) = (100, 100)$ and different $(m, p)$.
From Fig. \ref{fig: PE}, we can see that $1 - \mbb{P}(E)$ decreases rapidly for moderate values of $m$ (waterfall region), but almost stops decreasing for large $m$ (error floor region).

We note that in the waterfall (resp. error floor) region, the error patterns of incorrect received symbols have a relatively high (resp. low) probability to be linearly independent.
Accordingly, we can approximate $\mbb{P}(E)$ in the waterfall region based on the following theorem.

\begin{theorem}\label{theorem: P(E) > P(E1, E3)}
($E_1, E_3$) is a sufficient condition for $E$.
We have
\begin{align}\label{eqn: P(E) > P(E1, E3)}
    \mbb{P}(E) &\geq \mbb{P}(E_1, E_3)\nonumber
    \\&= \sum_{i = \max(n, m-l)}^m \binom{m}{i} p^i (1-p)^{m-i} \times \nonumber
    \\&\quad\quad\quad\quad\quad\quad\quad\quad P_{rk}(i, n) P_{rk}^*(l, m-i).
\end{align}
\end{theorem}

\begin{IEEEproof}
$E_3$ is a sufficient condition for $E_2$.
Then, ($E_1, E_3$) is a sufficient condition for $E$ according to Lemma \ref{lemma: E E1 E2}, leading to $\mbb{P}(E) \geq \mbb{P}(E_1, E_3)$.
Moreover, in \eqref{eqn: P(E) > P(E1, E3)}, $\binom{m}{i} p^i (1-p)^{m-i}$ is the probability that there exist $i$ correct received symbols.
Given this condition, $E_1$ and $E_3$ are independent.
Accordingly, $P_{rk}(i, n)$ (see Lemma \ref{lemma: rank property})  is the probability for $E_1$ to happen.
On the other hand, the $m-i$ incorrect received symbols form a $(m-i) \times (n+l)$ matrix in which each row is not a zero vector.
Hence, $P_{rk}^*(l, m-i)$ (see Lemma \ref{lemma: rank property no zero row}) is the probability for the matrix to have rank $m - i$ (i.e., for $E_3$ to happen).
This completes the proof.
\end{IEEEproof}

The time complexity for computing \eqref{eqn: P(E) > P(E1, E3)} is $O(\max(m, l))$, which is much lower than that for computing \eqref{eqn: P(E) = sum P_E}.
Note that as $m$ increases, $1 - \mbb{P}(E_1)$ keeps decreasing from 1 and $1 - \mbb{P}(E_3)$ keeps increasing from 0.
This explains why $1 - \mbb{P}(E_1, E_3)$ first  decreases and then increases, as can be seen from Fig. \ref{fig: PE}.
Moreover, we can see from Fig. \ref{fig: PE} that $\mbb{P}(E_1, E_3)$ coincides very well with $\mbb{P}(E)$ in almost the whole waterfall region.
This verifies that in the waterfall region,  $(E_1, E_3)$ (resp. $E_3$) is the major event for $E$ (resp. $E_2$).

We further approximate the computation in Theorem \ref{theorem: P(E) > P(E1, E3)} based on the following theorem.

\begin{theorem}\label{theorem: P(E) O(1)}
For any $\epsilon \in (0, p)$, let $n(\epsilon) = \lceil (p - \epsilon) m \rceil$.
If $n(\epsilon) > n$ and $m - n(\epsilon) < l$, we have
\begin{align}\label{eqn: P(E) O(1)}
    \mbb{P}(E) &\geq \mbb{P}(E_1, E_3)  \nonumber
    \\& \geq (1 - e^{-2 \epsilon^2 m})(1 - 2^{n - n(\epsilon)})  (1 - 2^{m - n(\epsilon) - l}).
\end{align}
\end{theorem}

\begin{IEEEproof}
The proof is similar to that of Theorem \ref{theorem: P(E) > P(E1, E3)}.
According to Lemma \ref{lemma: hoeffding inequality}, $1 - e^{-2 \epsilon^2 m}$ is the lower bound of the probability that there are at least $n(\epsilon)$ correct received symbols.
Given this condition, $E_1$ and $E_3$ are independent.
Moreover, $1 - 2^{n - n(\epsilon)}$ and $1 - 2^{m - n(\epsilon) - l}$ are the lower bounds of the probabilities for $E_1$ and $E_3$ to happen according to Lemmas \ref{lemma: rank property} and \ref{lemma: rank property no zero row},  respectively.
This completes the proof.
\end{IEEEproof}

The time complexity for computing \eqref{eqn: P(E) O(1)} is $O(1)$.
We remark that \eqref{eqn: m > n/p}, \eqref{eqn: l > n(1-p)/p}, and $m < l/(1 - p)$ are necessary for Theorem \ref{theorem: P(E) O(1)} to have a valid $\epsilon$.
The main purpose of Theorem \ref{theorem: P(E) O(1)} is to predict the behavior of $\mbb{P}(E)$ for large $(n, m, l)$ in which case it is impossible to compute $\mbb{P}(E)$ based on \eqref{eqn: P(E) = sum P_E} or \eqref{eqn: P(E) > P(E1, E3)}.
When $n, m,$ and $l$ are multiplied by a factor $\nu$ and $\nu \to \infty$, $\epsilon$ can be kept unchanged and  the lower bound given by \eqref{eqn: P(E) O(1)} can approach 1.
However, it seems hard to find the best $\epsilon$ so as to maximize the lower bound of \eqref{eqn: P(E) O(1)}.

\subsection{Analysis of $\mbb{P}(G)$}

We use $\mbb{P}(G)$ to approximate $\mbb{P}(E, F)$ due to two considerations.
First, $G = (E_1, E_3, F)$ is a sufficient condition for $(E, F)$ such that
\[
    \mbb{P}(E, F) \geq \mbb{P}(G).
\]
Second, inspired by Section \ref{subsection: P(E)},  it is shown that in the waterfall region (for moderate values of $m$), $\mbb{P}(E_1, E_3)$ coincides very well with $\mbb{P}(E)$, implying $\mbb{P}(G)$ should also coincide very well with $\mbb{P}(E, F)$ in the waterfall region.

Given $(p, n, l, m)$, we can explicitly compute $\mbb{P}(G)$ by using dynamic programming again  \cite[Section 15.3]{introAlgo01}.
To this end, for $0 \leq i \leq m$ and $0 \leq r_0 \leq r \leq n$, assuming $\mb{D}$ consists of $i$ arbitrary correct received symbols whose basis is $B(\mb{D})$, we define $P_G(i, r, r_0)$ as the probability that $|B(\mb{D})| = r$ and there are $r_0$ basis elements of $B(\mb{D})$ attending zero LRs of $\mb{D} \setminus B(\mb{D})$.
We have the following theorem.

\begin{theorem}\label{theorem: P(G) = sum P_G}
We have
\begin{align}\label{eqn: P(G) = sum P_G}
    \mbb{P}(G)&= \sum_{i = \max(n, m-l)}^m \binom{m}{i} p^i (1-p)^{m-i} \times \nonumber
    \\&\quad\quad\quad\quad\quad\quad\quad\quad P_{rk}^*(l, m-i) P_G(i, n, 0).
\end{align}
Moreover, we have $P_G(0, 0, 0) = 1$, and for $0 < i \leq m$ and $0 \leq r_0 \leq r \leq n$, we have
\begin{align}\label{eqn: P_G}
    P_G(i, r, r_0) = &P_G(i - 1, r-1, r_0 - 1) (1 - 2^{r-1-n}) +\nonumber
    \\& \sum_{r_0 \leq r'_0 \leq r} P_G(i - 1, r, r'_0) \binom{r'_0}{r_0} 2^{r - r'_0-n},
\end{align}
where for simplicity, we let $P_G(\cdot, r', r'_0) = 0$ for any $r' < 0$ or $r'_0 < 0$.
\end{theorem}

\begin{IEEEproof}
See Appendix \ref{appendix: P(G) = sum P_G}.
\end{IEEEproof}

\begin{figure}[t]
\centering
\includegraphics[scale = 0.5]{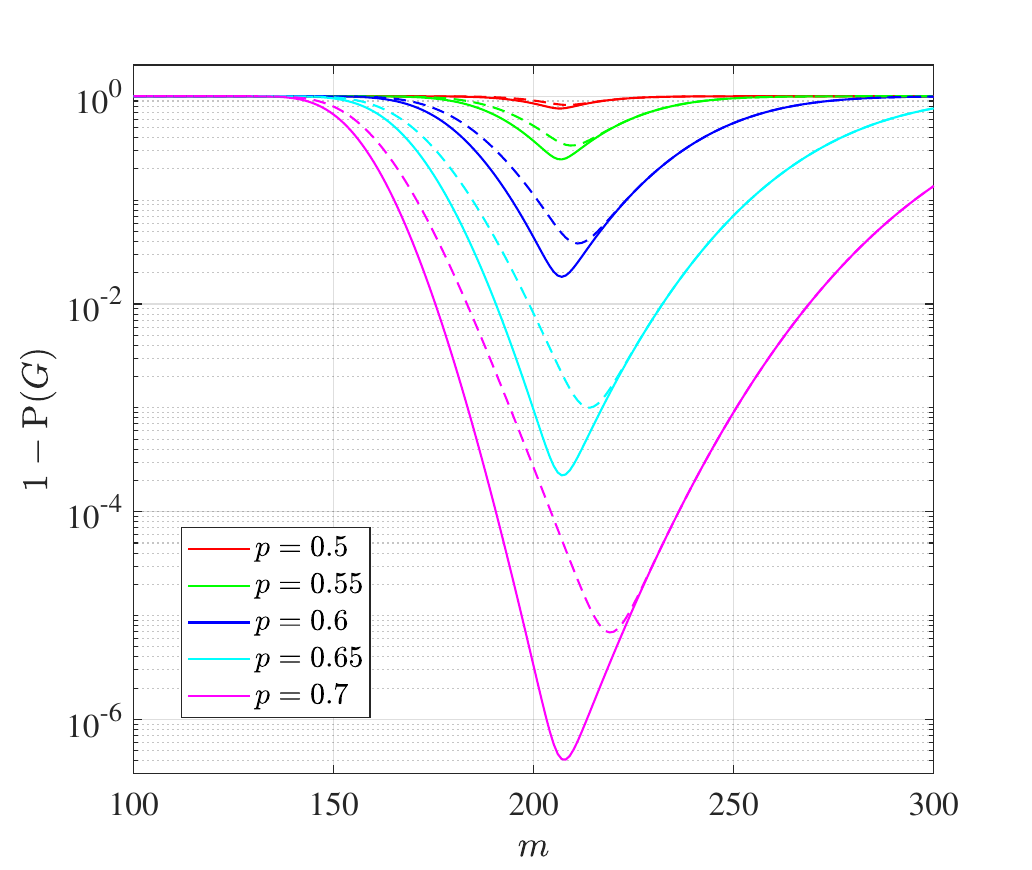}
\caption{Numerical result of $1 - \mbb{P}(G)$ for $(n, l) = (100, 100)$, where solid and dash lines are computed based on \eqref{eqn: P(G) = sum P_G} and \eqref{eqn: P(G) = sum P_G approximate}, respectively.}
\label{fig: PG}
\end{figure}

According to Theorem \ref{theorem: P(G) = sum P_G}, we can compute $P_G(i, r, r_0)$ for all $0 \leq i \leq m$ and $0 \leq r_0 \leq r \leq n$ with time complexity $O(mn^3)$, and then compute $\mbb{P}(G)$ with time complexity $O(\max(m, l))$.
As an example, we show the numerical result of $1 - \mbb{P}(G)$ in Fig. \ref{fig: PG} for $(n, l) = (100, 100)$ and different $(m, p)$.
From Fig. \ref{fig: PG}, we can see that $1 - \mbb{P}(G)$ first decreases and then increases as $m$ increases.
This trend of $1-\mbb{P}(G)$ coincides well with that of $1-\mbb{P}(E_1, E_3)$ shown by Fig. \ref{fig: PE}.

We are now to approximate $P_G(i, n, 0)$ so as to reduce the complexity for computing $\mbb{P}(G)$.

\begin{lemma}\label{lemma: P_G approximate}
For $i \geq n$, we have
\begin{align}
    P_G(i, n, 0) \geq &\max_{0 \leq h \leq i-n} P_{rk}(i-h, n) (1 - 2^{-h})^n \label{eqn: P_G approximate 1}
    \\\geq& \max_{0 \leq h \leq i-n} (1 - 2^{n-i+h}) (1 - 2^{-h})^n \label{eqn: P_G approximate 2}
    \\=& \max_{\lfloor \hat{h} \rfloor \leq h \leq \lceil \hat{h} \rceil } (1 - 2^{n-i+h}) (1 - 2^{-h})^n, \label{eqn: P_G approximate 3}
\end{align}
where $\hat{h} = \log_2\left( \sqrt{(n-1)^2 + n2^{i-n+2}} - n + 1 \right) - 1$.
\end{lemma}

\begin{IEEEproof}
For $i \geq n$ and given $h \in \{0, 1, \ldots, i-n\}$, denote $B$ as the basis of $i-h$ arbitrary correct received symbols.
$P_{rk}(i-h, n)$ (see Lemma \ref{lemma: rank property}) is the probability for $|B| = n$.
Further given $|B| = n$, all basis elements of $B$ have probability $1/2$ to independently attend the LR of an arbitrary correct received symbol.
Thus, among the LRs of $h$ arbitrary correct received symbols, $1 - 2^{-h}$ is the probability for a basis element of $B$ to attend at least one LR, and $(1 - 2^{-h})^n$ is the probability for all the $n$ basis elements of $B$  to attend at least one LR.
This completes the proof of \eqref{eqn: P_G approximate 1}.
According to Lemma \ref{lemma: rank property}, \eqref{eqn: P_G approximate 2} holds.
Moreover, define $f(h) = (1 - 2^{n-i+h}) (1 - 2^{-h})^n$ as a function of $h$ on the real interval $[0, i-n]$.
By computing $\frac{\textrm{d} f}{\textrm{d} h}$, we can easily see that $f(h)$ increases in $[0, \hat{h}]$ and decreases in $[\hat{h}, i-n]$, leading to  \eqref{eqn: P_G approximate 3}.
\end{IEEEproof}

\begin{theorem}\label{theorem: P(G) = sum P_G approximate}
We have
\begin{align}\label{eqn: P(G) = sum P_G approximate}
    \mbb{P}(G)\geq& \sum_{i = \max(n, m-l)}^m \binom{m}{i} p^i (1-p)^{m-i} P_{rk}^*(l, m-i) \times \nonumber
    \\& \quad \max_{\lfloor \hat{h} \rfloor \leq h \leq \lceil \hat{h} \rceil } P_{rk}(i-h, n) (1 - 2^{-h})^n,
\end{align}
where $\hat{h} = \log_2\left( \sqrt{(n-1)^2 + n2^{i-n+2}} - n + 1 \right) - 1$.
\end{theorem}

\begin{IEEEproof}
This theorem is a combined result of Theorem \ref{theorem: P(G) = sum P_G} and Lemma \ref{lemma: P_G approximate}.
\end{IEEEproof}

The time complexity for computing \eqref{eqn: P(G) = sum P_G approximate} is $O(\max(m, l))$, which is much lower than that for computing \eqref{eqn: P(G) = sum P_G}.
From Fig. \ref{fig: PG}, we can see that \eqref{eqn: P(G) = sum P_G approximate} offers an acceptable lower bound for $\mbb{P}(G)$.

We further approximate the computation in Theorem \ref{theorem: P(G) = sum P_G approximate} based on the following theorem.

\begin{theorem}\label{theorem: P(G) O(1)}
For any $\epsilon \in (0, p)$, let $n(\epsilon) = \lceil (p - \epsilon) m \rceil$.
If $n(\epsilon) > n$ and $m - n(\epsilon) < l$, we have
\begin{align}\label{eqn: P(G) O(1)}
    \mbb{P}(G)\geq& (1 - e^{-2 \epsilon^2 m})(1 - 2^{m - n(\epsilon) - l}) \times  \nonumber
    \\&  \max_{\lfloor \hat{h} \rfloor \leq h \leq \lceil \hat{h} \rceil } (1 - 2^{n-n(\epsilon)+h}) (1 - 2^{-h})^n,
\end{align}
where $\hat{h} = \log_2\left( \sqrt{(n-1)^2 + n2^{n(\epsilon)-n+2}} - n + 1 \right) - 1$.
\end{theorem}

\begin{IEEEproof}
According to Lemma \ref{lemma: hoeffding inequality}, $1 - e^{-2 \epsilon^2 m}$ is the lower bound of the probability that there are at least $n(\epsilon)$ correct received symbols.
Then, we can easily derive \eqref{eqn: P(G) O(1)} by using $1 - e^{-2 \epsilon^2 m}$ and $n(\epsilon)$ to replace $\sum_{i = \max(n, m-l)}^m \binom{m}{i} p^i (1-p)^{m-i}$ and $i$ in \eqref{eqn: P(G) = sum P_G approximate}, respectively.
\end{IEEEproof}

The time complexity for computing \eqref{eqn: P(G) O(1)} is $O(1)$.
The main purpose of Theorem \ref{theorem: P(G) O(1)} is to predict the behavior of $\mbb{P}(G)$ for large $(n, m, l)$ in which case it is impossible to compute $\mbb{P}(G)$ based on \eqref{eqn: P(G) = sum P_G} or \eqref{eqn: P(G) = sum P_G approximate}.
When $n, m,$ and $l$ are multiplied by a factor $\nu$ and $\nu \to \infty$, $\epsilon$ can be kept unchanged; meanwhile, the lower bound given by \eqref{eqn: P(G) O(1)} can approach 1 (it is easier to see this for $h = (n(\epsilon) - n)/2$).
However, it seems hard to find the best $\epsilon$ so as to maximize the lower bound of \eqref{eqn: P(G) O(1)}.

We now connect Theorem \ref{theorem: P(G) O(1)} to $C_{\text{erasure}}$ given by \eqref{eqn: c erasure}.
For given $\epsilon \in (0, p)$, let $m = n(1+\epsilon)/(p-\epsilon)$ and $l = m - n(\epsilon) + \epsilon n$.
We can see that $\mbb{P}(G) \to 1$ as $n \to \infty$ according to Theorem \ref{theorem: P(G) O(1)}.
Meanwhile, the overall transmission rate is given by $nl / m = l(p - \epsilon)/(1 + \epsilon)$, which can be made as close to $pl$ as needed since $\epsilon$ can be set arbitrarily small.
On the other hand, using $\log_2(m)$ bits to denote the seed of an encoded symbol is sufficient.
Then, a total of $l+\log_2(m)$ bits need to be transmitted for an encoded symbol.
As a result, $w$ in \eqref{eqn: c erasure} is equal to $l + \log_2(m)$, leading to $C_{\text{erasure}} \sim p l$.
This indicates that $C_{\text{erasure}}$ is achieved in this case.

\subsection{Analysis of $\mbb{P}(E, F)$ for Relatively Large $m$}\label{subsection: Prob for large m}

For fixed $(p, n, l)$, as $m$ increases, $E_3$ becomes less likely to happen than $(E, F)$ but $E_3$ is actually not necessary for $(E, F)$.
Thus, it is not suitable to use $\mbb{P}(G)$ to approximate $\mbb{P}(E, F)$ for relatively large $m$.
On the other hand, for relatively large $m$, $1-\mbb{P}(E)$ almost stops decreasing (encountering error floor), as can be seen from Fig. \ref{fig: PE}.
Since $\mbb{P}(E)$ is an upper bound of $\mbb{P}(E, F)$,  $1 - \mbb{P}(E, F)$ must also encounter the error floor.
To see whether $1 - \mbb{P}(E, F)$ can approach $1 - \mbb{P}(E)$ or not, we need to check how $\mbb{P}(F | E)$ behaves as $m$ continuously increases.

\begin{theorem}\label{theorem: probability for attending LR}
Let $B_r$ be an arbitrary  set of $r$ linearly independent received symbols, which includes $n$ correct and $r - n$ incorrect received symbols ($n \leq r \leq n+l$).
Let $(\mb{a}, \mb{y})$ be an arbitrary received symbol which is an LC of $B_r$.
Define $p_c$ (resp. $p_e$) as the probability that a correct (resp.  incorrect) received symbol of $B_r$ attends the LR of $(\mb{a}, \mb{y})$.
We have
\begin{align*}
    &p_c = 1/2 \text{~and~}
    \\&p_e = \frac{(1 - p) 2^{r-n-1}/(2^{l} - 1)}{p + (1 - p) (2^{r-n} - 1)/(2^{l} - 1)} < p_c.
\end{align*}
\end{theorem}

\begin{IEEEproof}
See Appendix \ref{appendix: probability for attending LR}.
\end{IEEEproof}

Theorem \ref{theorem: probability for attending LR} indicates that if $E$ happens, the correct basis elements have a better chance to attend more LRs of $(\mb{A}, \mb{Y}) \setminus B(\mb{A}, \mb{Y})$ than the incorrect basis elements.
Combining with Lemma \ref{lemma: hoeffding inequality}, we are likely to have $\mbb{P}(F | E) \to 1$ and $\mbb{P}(E, F) \to \mbb{P}(E)$ as $m \to \infty$.
However, it seems not easy to get a satisfied bound of $\mbb{P}(E, F)$ for relatively large $m$, mainly because the incorrect basis elements do not independently attend the LRs of $(\mb{A}, \mb{Y}) \setminus B(\mb{A}, \mb{Y})$.

\subsection{Remarks}

According to the analysis in this section, it is clear that the error-correction performance of the BFA is determined by the quality of a  basis: Whether the basis contains $n$ correct basis elements (whether $E$ happens) and whether the correct basis elements attend more LRs   than the incorrect ones (whether $F$ happens).
We know that for given $(n, m, p)$, increasing $l$ will lead to lower FERs, since the incorrect basis elements are expected to attend fewer LRs or even none of the LRs.
On the other hand, for given $(n, l, p)$, as $m$ increases, the FER of BFA must first decrease for small $m$ and finally decrease to the error floor for (very) large $m$.
However,  for moderate $m$ where the least (resp. largest) number of LRs attended by correct (resp. incorrect) basis elements is relatively low (but not zero), the FER may decrease or increase depends on the quality of the basis.
This is observed in the simulations for practical LT codes  in Section \ref{section: simulation}, and we will given more explanations there.

\section{Simulation Results}\label{section: simulation}

\begin{figure*}[!t]
\centering

\subfigure[Straightforward implementation.]{%
  \includegraphics[scale = 0.5]{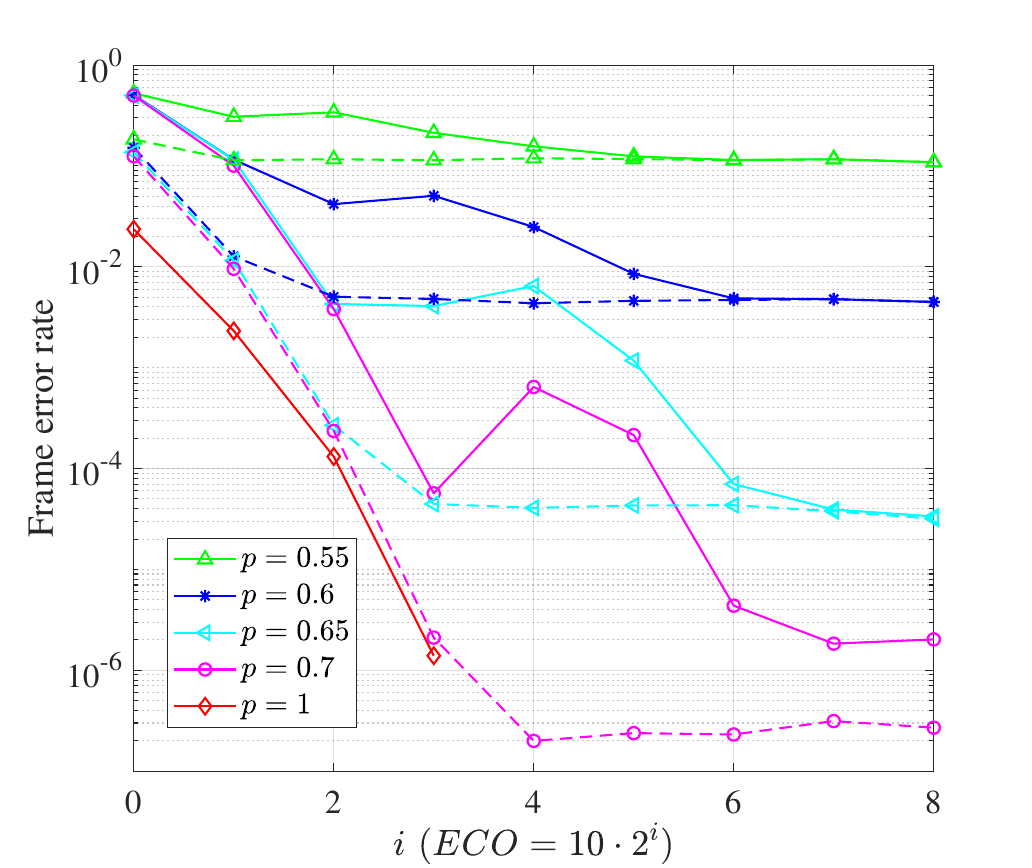}
}
\subfigure[Triangulation-based implementation.]{%
  \includegraphics[scale = 0.5]{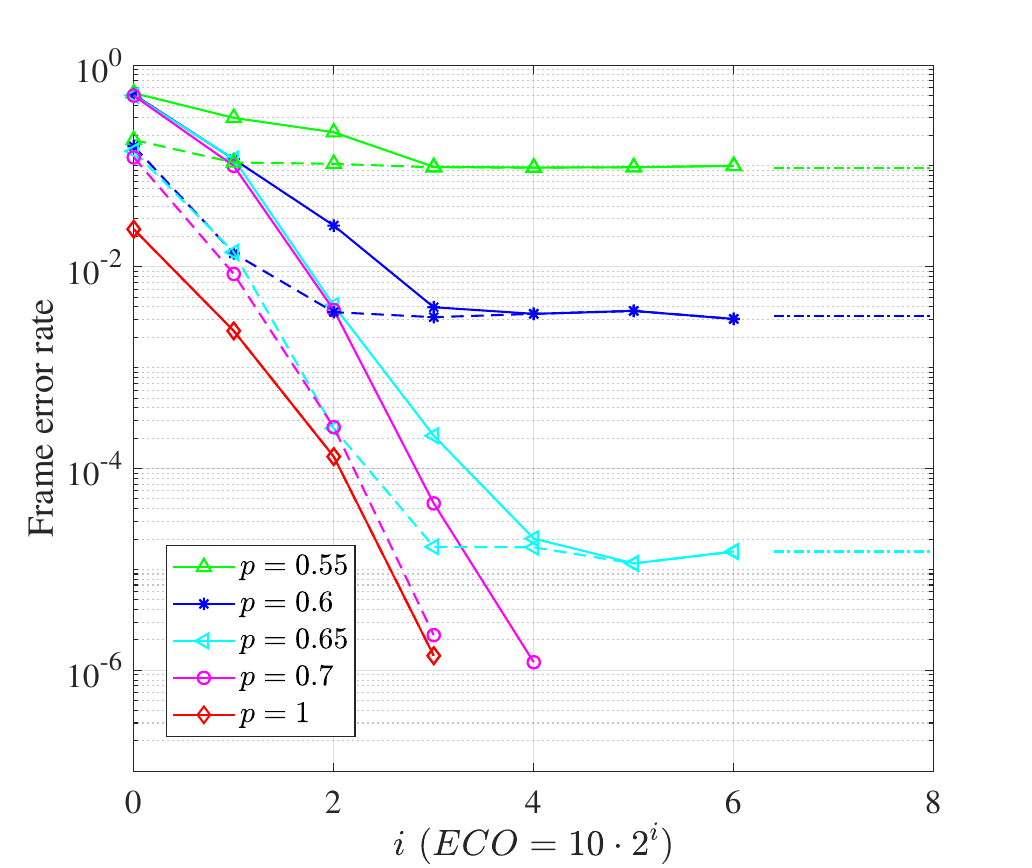}
}

\subfigure[Sorted-weight  implementation.]{%
  \includegraphics[scale = 0.5]{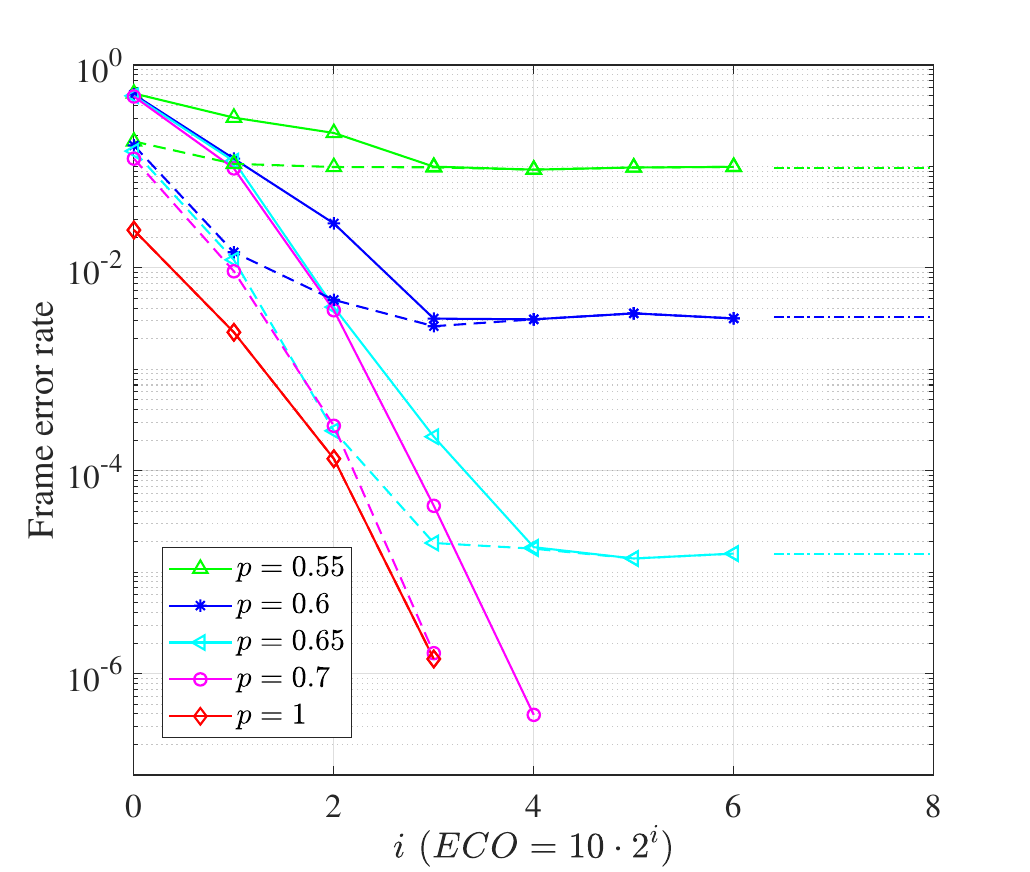}
}
\subfigure[BP algorithm\cite{etesami2006raptor}.]{%
  \includegraphics[scale = 0.5]{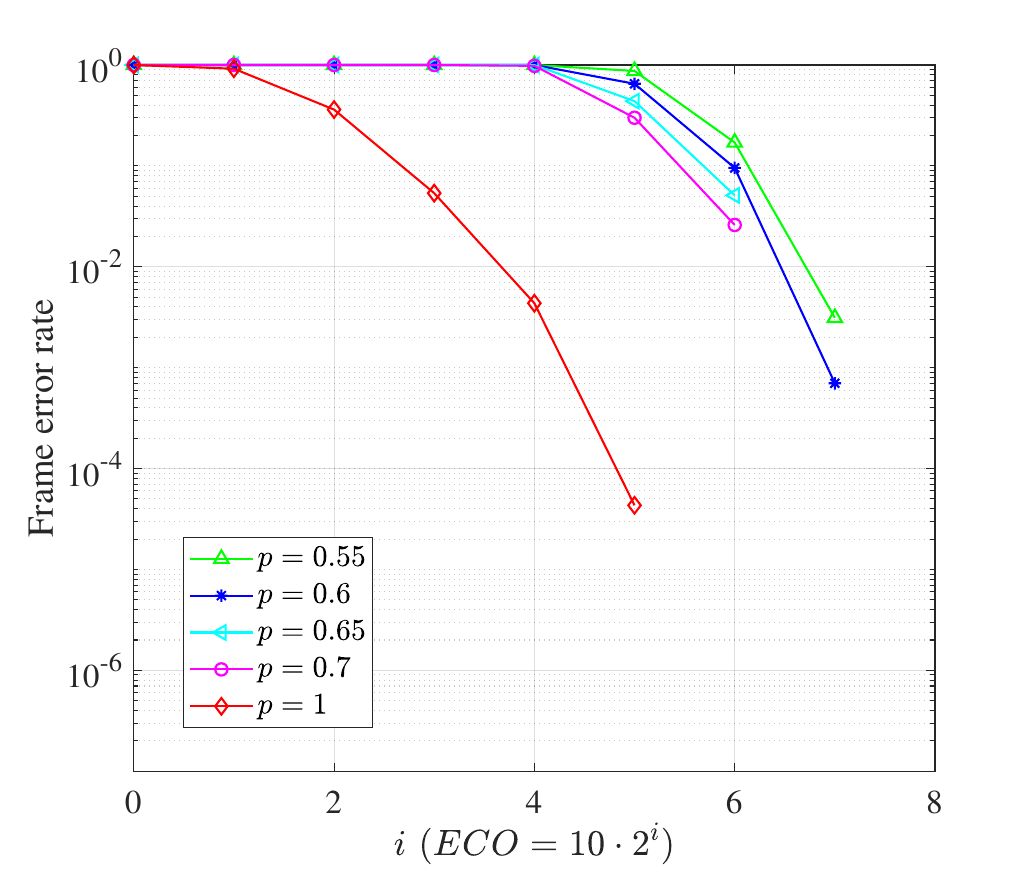}
}

\caption{Error-correction performance of the BFA with different implementations and of the BP algorithm \cite{etesami2006raptor} with 100 iterations.
We set $(n, l) = (100, 100)$ and use the RSD with $(\delta, c) = (0.01, 0.02)$.
For (a)--(c), solid (resp. dash) lines correspond to the frequencies that $(E, F)$ (resp. $E$) does not happen, and dash-dot lines are $1 - \mbb{P}(E)$ computed based on  \eqref{eqn: P(E) = sum P_E}. (The events $E$ and $F$ are defined in Section \ref{section: performance analysis}.)}
\label{fig: FER_l100_n100}
\end{figure*}

In this section, we present the simulation results for our proposed BFA and the BP algorithm \cite{etesami2006raptor}.
We realize the BFA using the three proposed implementations.
LT codes are employed, and we will specify the distributions used for generating the weights of LT encoded symbols for different simulation scenarios.
For any transmitted encoded symbol $(\mb{a}, \mb{y}) \in (\mbb{F}_2^{n}, \mbb{F}_2^l)$, it remains the same with probability $p$ at the receiver side, and with probability $1 - p$, an error pattern is uniformly selected from $\mbb{F}_2^l \setminus \{0^l\}$ to add into $\mb{y}$.
As a result, the BP algorithm uses the probability $p_b$ given by \eqref{eqn: pb} to compute the soft information of each data payload bit of a received symbol.

We present the error-correction performance of a decoding algorithm in the way of FER versus the expected correct overhead (ECO).
The FER corresponds to the frequency that the source symbols are not fully recovered. (In the BFA, the FER corresponds to the frequency that the event $(E, F)$  does not happen according to Lemma \ref{lemma: (E, F)}).
The ECO is defined by
\[
    ECO = p m - n,
\]
which is the  number of redundant symbols that are expected to be received correctly.
We present the error-correction performance in terms of FER versus ECO since the BFA needs to receive at least $n$ correct symbols for possible successful decoding, i.e., $ECO > 0$ is necessary for low FER.
Moreover, for $p = 1$, the definition of ECO is consistent with the definition of the overhead for erasure channels, and the BFA has the same error-correction performance as the inactivation decoding \cite{odlyzko1984discrete, lamacchia1991solving, he2020disjoint, lazaro2017fountain,  3GPP06, shokrollahi2005systems, shokrollahi2011raptor}.
Thus, we use the error-correction performance under $p = 1$ as a benchmark, and show how the performance changes for $p < 1$ under the same ECO.
We collect more than 50 frame errors for each simulation point.

In the first simulation scenario, we set $(n, l) = (100, 100)$ and generate LT encoded symbols using the RSD with $(\delta, c) = (0.01, 0.02)$ (see Definition \ref{def: ISD/RSD}).
We show the error-correction performance of the straightforward implementation, triangulation-based implementation,  sorted-weight implementation, and the BP algorithm \cite{etesami2006raptor} in Figs. \ref{fig: FER_l100_n100}(a)--(d), respectively.
We can see that, (i) the three implementations of the BFA can  significantly outperform the BP algorithm, except for cases where $p < 1$ and the ECO is extremely large.
For those cases, the BFA can have higher error floor than the BP algorithm.
(ii) The triangulation-based implementation and sorted-weight implementation perform very closely,  except that the sorted-weight implementation performs slightly better at $(p, ECO) = (0.7, 10 \cdot 2^4)$.
It is worth mentioning that both the triangulation-based implementation and the sorted-weight implementation have an error floor coinciding with the lower bound $1 - \mbb{P}(E)$ (see \eqref{eqn: P(E) = sum P_E}) that is derived for random fountain codes.
(iii) For relatively small ECOs (e.g., $ECO \leq 10 \cdot 2^2$ for $p = 0.6$ and $ECO \leq 10 \cdot 2^3$ for $p = 0.7$) where the incorrect basis elements are likely to attend zero LR, the straightforward implementation has a similar FER as the triangulation-based implementation (as well as the sorted-weight implementation).
However, as the ECO increases such that the incorrect basis elements starts to attend more and more LRs, the FER of the straightforward implementation first increases and finally re-decreases until encountering the error floor.

\begin{figure}[!t]
\centering
\includegraphics[scale = 0.5]{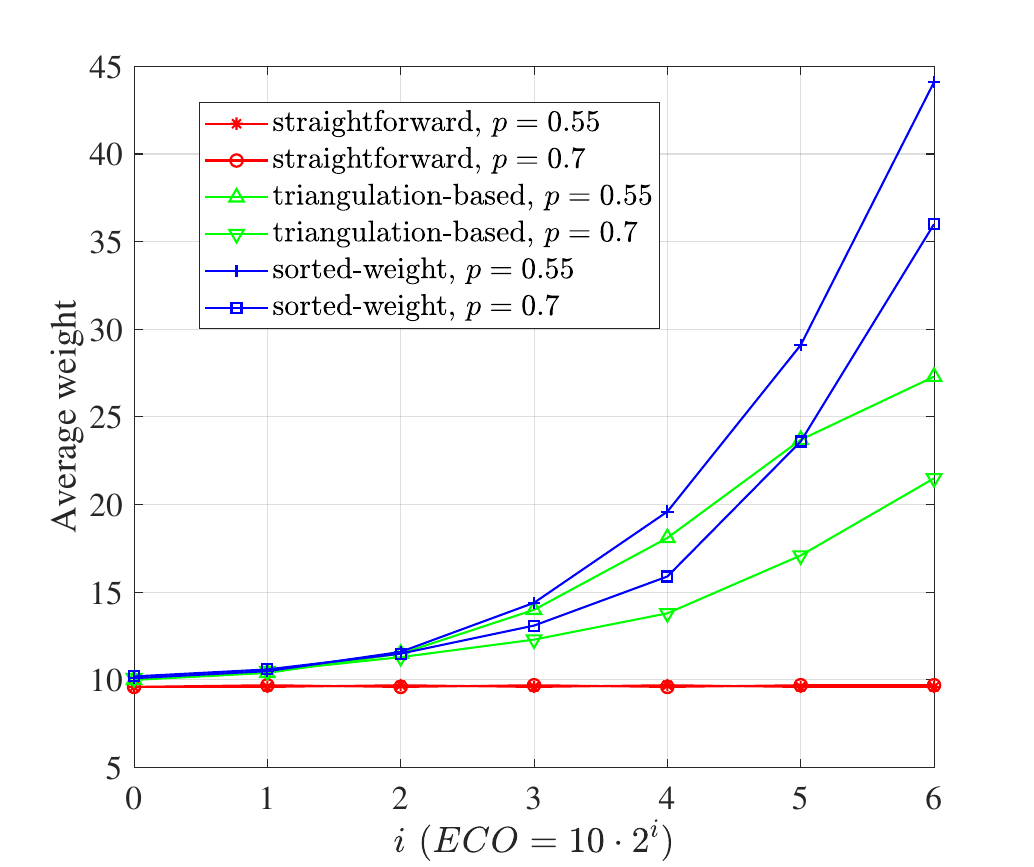}
\caption{Average weights of basis elements for the straightforward, triangulation-based, and sorted-weight implementations of the BFA. }
\label{fig: FER_l100_n100_weight}
\end{figure}

It is clear that the three implementations of the BFA generate a different basis, which is the reason for having different error-correction performance.
More specifically, the average weights of basis elements for different implementations are shown in Fig. \ref{fig: FER_l100_n100_weight}, where we only present the results for $p = 0.55$ and $p = 0.7$ to avoid including too many curves  in Fig. \ref{fig: FER_l100_n100_weight} (the average weights  for $p = 0.6$ and $p = 0.65$ have a similar trend).
We can see that for a given $p$, the straightforward implementation, triangulation-based implementation, and sorted-weight implementation have increasing average weights.
Combining with the results presented in Fig. \ref{fig: FER_l100_n100}, it is suggested that a basis with larger average weight of its basis elements generally has better error-correction performance, as previously stated in Observation 1.
We guess that this phenomenon is caused by the difference between how the sparse and dense encoded symbols affect the rank properties, as discussed in \cite{shokrollahi2011raptor}.
It is found in \cite{shokrollahi2011raptor} that for practical LT codes which have a sparse $\mb{A}$, adding a few dense rows into $\mb{A}$ (increasing the average weight of encoded symbols) can mimic the behavior of the random fountain codes and then improve the error-correction performance.

%
%
%
%
%

\begin{figure}[!t]
\centering

\subfigure[RSD with $(\delta, c) = (0.01, 0.02)$.]{%
  \includegraphics[scale = 0.5]{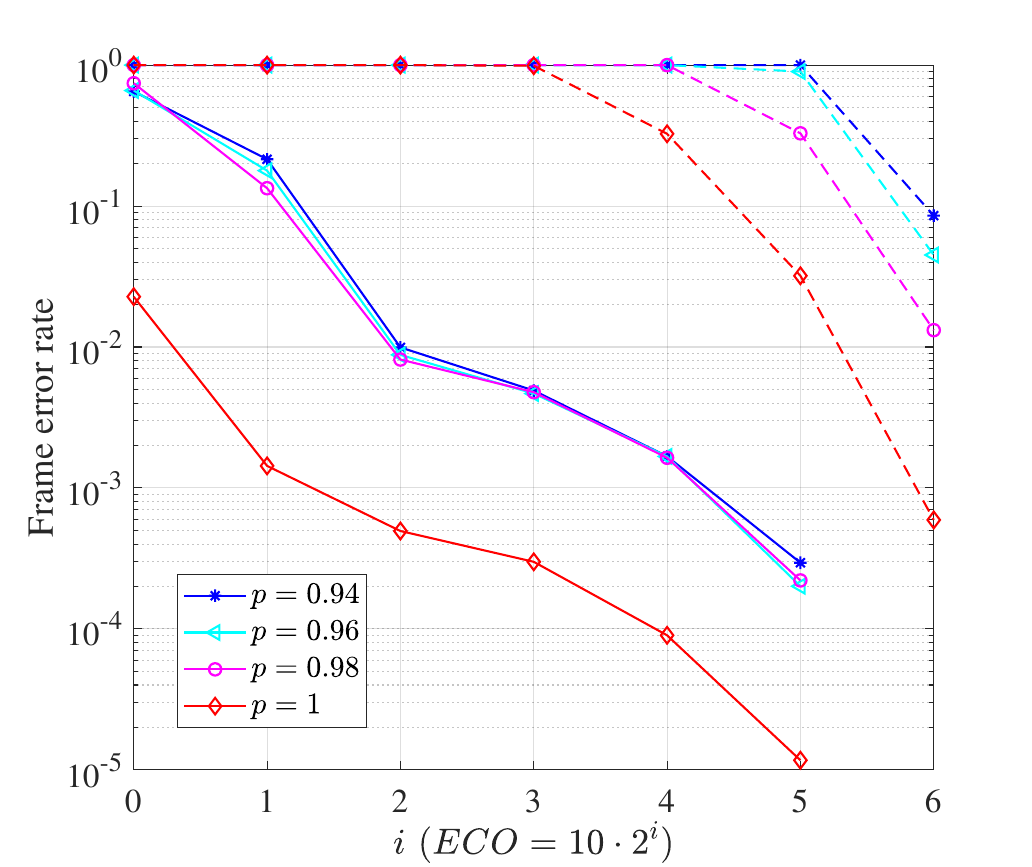}
}
\subfigure[Weight distribution of \eqref{eqn: omega(x)}.]{%
  \includegraphics[scale = 0.5]{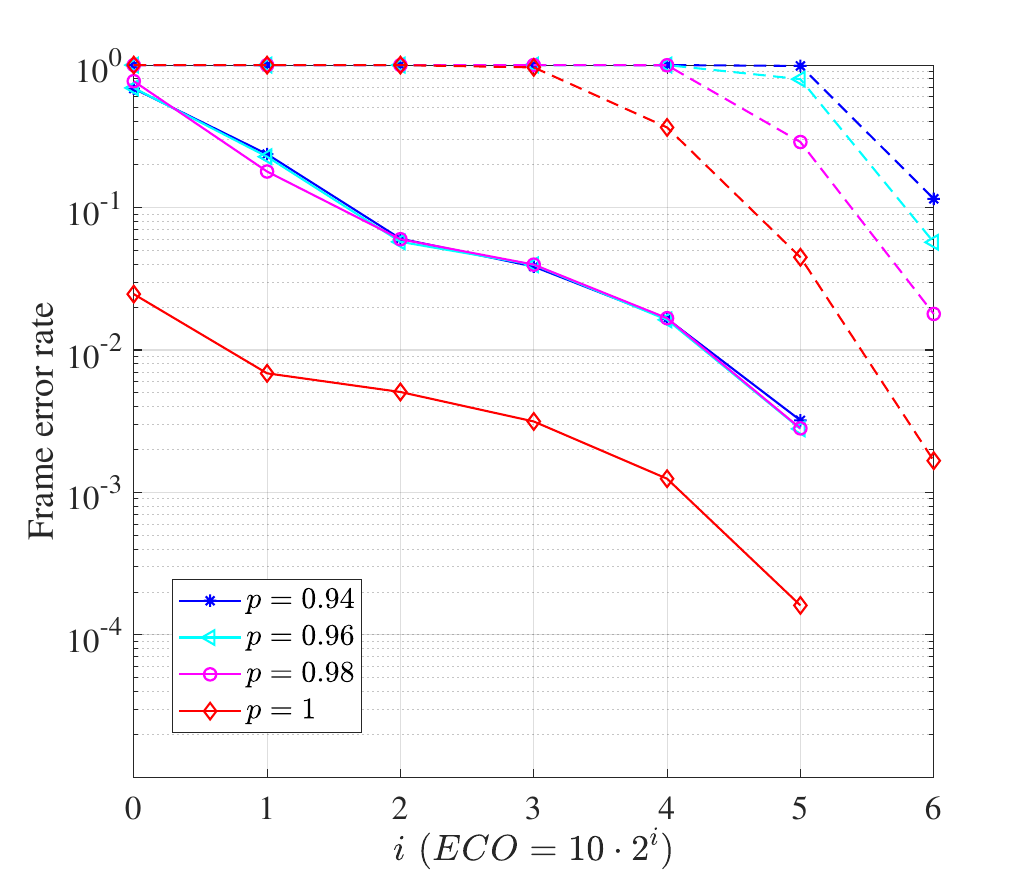}
}

\caption{Error-correction performance of the BFA (solid lines) with triangulation-based implementation and of the BP algorithm \cite{etesami2006raptor} (dash lines) with 100 iterations.
We set $(n, l) = (1000, 100)$.}
\label{fig: FER_l100_n1000}
\end{figure}

In the second scenario, we set $(n, l) = (1000, 100)$.
Two weight distributions are used for generating LT encoded symbols.
One is the RSD with $(\delta, c) = (0.01, 0.02)$.
The other is
\begin{align}\label{eqn: omega(x)}
\Omega(x) = \,&0.006x + 0.492x^2 + 0.0339x^3 + 0.2403x^4 \nonumber
    \\&+ 0.006x^5 + 0.095x^8 + 0.049x^{14} + 0.018x^{30} \nonumber
    \\&+ 0.0356x^{33} + 0.033x^{200},
\end{align}
where the coefficient of $x^i$ is the probability for choosing weight $i$.
$\Omega(x)$ is optimized in \cite{etesami2006raptor} for the BP decoding of LT codes over AWGN channel.
The error-correction performance of the BFA with triangulation-based implementation and of the BP algorithm is shown in Fig. \ref{fig: FER_l100_n1000}.
We can see that the BFA significantly outperforms the BP algorithm.
We remark that we ignore the error-correction performance of the straightforward implementation and sorted-weight implementation since they perform comparably with the triangulation-based implementation for $ECO \leq 10 \cdot 2^5$ and $p \in \{0.94, 0.96, 0.98, 1\}$.
In fact, in this region, the expected number of incorrect received symbols, given by $(ECO + n) (1 - p) / p$, is at most $84.3$.
Noting that $l = 100$, it indicates that the incorrect received symbols have a good chance to be linearly independent (attend zero LR).

\begin{figure}[!t]
\centering
\includegraphics[scale = 0.5]{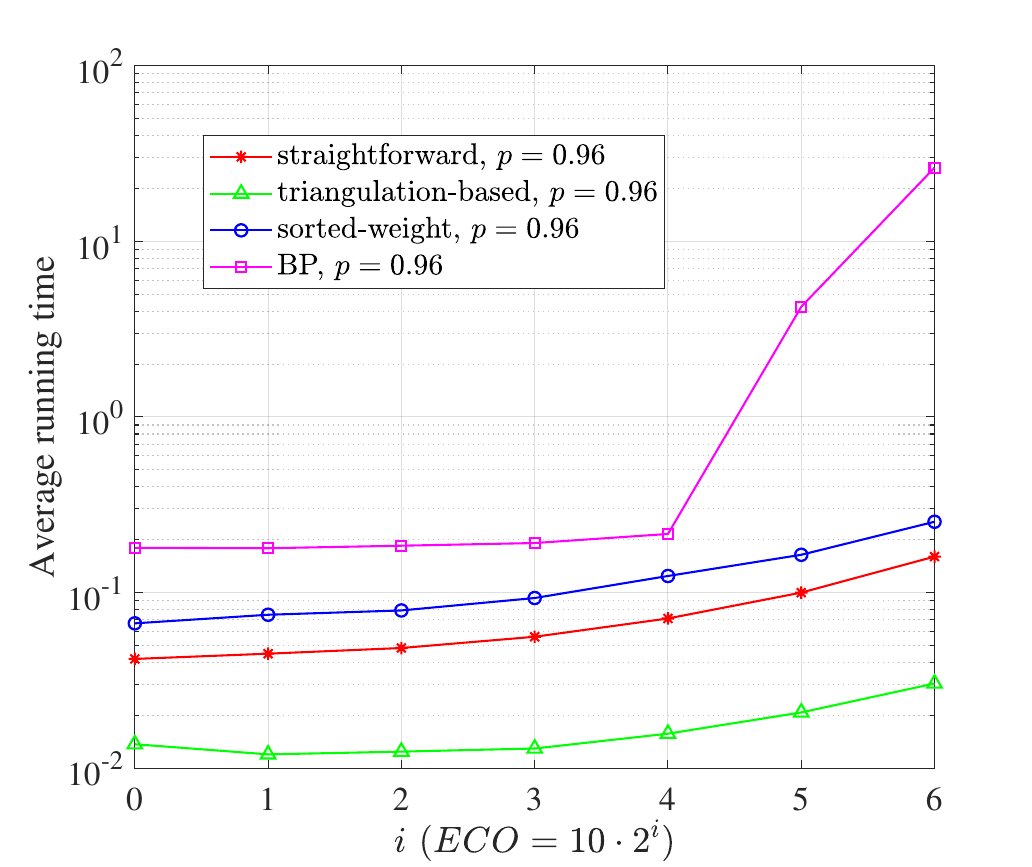}
\caption{Average running time of the BFA with different implementations and of the BP algorithm \cite{etesami2006raptor} with 100 iterations. }
\label{fig: FER_l100_n1000_time}
\end{figure}

The complexity of different decoding algorithms plays an import role in practice.
Our simulation results show that the triangulation-based implementation, straightforward implementation, sorted-weight implementation, and the BP algorithm run from the fastest to the slowest, respectively.
In particular, the triangulation-based implementation generally runs tens to hundreds of times faster than the BP algorithm.
As an example, for $p = 0.96$ and the RSD with $(\delta, c) = (0.01, 0.02)$, we illustrate the average running time of different decoding algorithms in Fig. \ref{fig: FER_l100_n1000_time}.
We think that one major reason leading to the big gap between the running time of the BFA and the BP algorithm is because the BP algorithm has many complicated floating-point operations, while the BFA only has simple integer operations and many of them are bit-XOR operations.
However, there is no guarantee that this is a general behaviour, since the running time also depends on the actual implementation.
For easy reference, we  share our C++ program used for all the simulations involved in this paper at https://github.com/Hestandy/Basis-Finding.


\section{Conclusions and Remarks}\label{section: conclusion}

In this paper, we have proposed a basis-finding algorithm (BFA) for decoding fountain codes  where the received symbols may have substitution errors.
It is motivated by the application of fountain codes in DNA-based data storage systems where the inner code decoding, which generally has undetectable errors, is performed before the outer fountain code decoding.
The BFA only needs to take the received symbols as input, but it can also make use of the reliability information (if available) of the received symbols for decoding.
We have developed three implementations for the BFA, namely the straightforward, triangulation-based, and sorted-weight implementations.
They are based on Gaussian elimination  and thus have polynomial time complexities.
Moreover, they generally have increasing average weights of basis elements and decreasing frame error rates (FERs).

Extensive simulations with Luby transform (LT) codes have revealed that, the BFA (with any of the three implementations) can have significantly lower FER than the belief propagation (BP) algorithm \cite{etesami2006raptor}, except that the BFA can have higher error floor when $p < 1$ and the expected correct overhead (ECO) is extremely large.
We remark that for the case with $p = 1$ which corresponds to an erasure channel, the BFA is essentially the same as the inactivation decoding \cite{odlyzko1984discrete, lamacchia1991solving, he2020disjoint, lazaro2017fountain,  3GPP06, shokrollahi2005systems, shokrollahi2011raptor}, and it is well known that the inactivation decoding can significantly outperform the BP algorithm for any given ECO.
On the other hand, in the DNA-based data storage systems, the ECO is predefined and fixed before writing information into DNA strands which correspond to codewords of the inner codes/encoded symbols of the outer fountain codes.
As a result, given that a valid DNA strand (i.e., being a codeword of the inner codes and a correct encoded symbol of the outer fountain codes) is very unlikely to change into another valid DNA strand, the maximum possible number of distinct valid DNA strands that can be recovered after the inner code decoding can still be considered as fixed (determined by the predefined ECO).
Moreover, the predefined ECO is generally small in order to ensure a high net information density.
This implies that when applied to DNA-based data storage systems, (i) the BFA may always have a much lower FER than the BP algorithm;
(ii) the triangulation-based implementation is more preferable due to having the lowest complexity and the same best FER among the three implementations.

For random fountain codes, we have derived the theoretical bounds for the FER of the BFA.
The theoretical results can accurately predict the error floor of the BFA with triangulation-based and sorted-weight implementations, according to the simulation results.
Meanwhile, theoretical analysis indicates that the error floor of the BFA is mainly due to the existence of the incorrect received symbols such that the basis of the received symbols may not contain a sufficient number of correct received symbols.
One possible way to reduce the error floor is to make use of the reliability information (if available) of the received symbols when generating the basis, such as making use of the number of occurrence (proportional to the reliability) of each received symbol in the DNA-based data storage systems.

%

\appendices

\section{Proof of Theorem \ref{theorem: P(E) = sum P_E}}\label{appendix: P(E) = sum P_E}

Based on the definition of $P_E(\cdot, \cdot, \cdot)$, it is easy to see that \eqref{eqn: P(E) = sum P_E} holds and $P_E(0, 0, 0) = 1$.
We now compute $P_E(i, n_c, n_e)$ for $0 < i \leq m, 0 \leq |B_c(i)| = n_c \leq n$, and $0 \leq |B_e(i)| = n_e \leq l$.
We only need to compute how $P_E(i-1, \cdot, \cdot)$ contributes to $P_E(i, n_c, n_e)$ by taking into account all the situations that happen to the $i$-th received symbol $(\mb{a}_i, \mb{y}_i)$.
According to the first two assumptions at the beginning of Section \ref{section: performance analysis}, if $(\mb{a}_i, \mb{y}_i)$ is correct, it has $2^n$ equally like possibilities to occur; otherwise, it has $2^{n+l} - 2^n$ equally like possibilities to occur.
Moreover, according to the third assumption at the beginning of Section \ref{section: performance analysis}, $(\mb{a}_i, \mb{y}_i) \in B(i)$ if and only if (iff) $(\mb{a}_i, \mb{y}_i)$ is not an LC of $B(i-1)$.

\emph{Case 1 ($(\mb{a}_i, \mb{y}_i)$ is correct and $(\mb{a}_i, \mb{y}_i) \notin B(i)$):} This indicates a way that $P_E(i-1, n_c, n_e)$ contributes to $P_E(i, n_c, n_e)$.
In this case, any basis element of $B_e(i)$  cannot attend the LR of $(\mb{a}_i, \mb{y}_i)$, since the error pattern of $(\mb{a}_i, \mb{y}_i)$ is a zero vector and the error patterns of $B_e(i)$ are linearly independent.
Moreover, as $(\mb{a}_i, \mb{y}_i) \notin B(i)$, $(\mb{a}_i, \mb{y}_i)$ must be an LC of $B_c(i - 1)$ and thus has $2^{n_c}$ possibilities to occur.
Therefore, Case 1 happens with probability $p 2^{n_c}/2^n$.

\emph{Case 2 ($(\mb{a}_i, \mb{y}_i)$ is incorrect and $(\mb{a}_i, \mb{y}_i) \notin B(i)$):} This indicates the other way that $P_E(i-1, n_c, n_e)$ contributes to $P_E(i, n_c, n_e)$.
In this case, $(\mb{a}_i, \mb{y}_i)$ is an LC of $B(i-1)$ and has $2^{n_c + n_e} - 2^{n_c}$ possibilities to occur.
Thus, Case 2 happens with probability $(1 - p) (2^{n_c + n_e} - 2^{n_c})/(2^{n+l} - 2^n)$.

\emph{Case 3 ($(\mb{a}_i, \mb{y}_i)$ is correct and $(\mb{a}_i, \mb{y}_i) \in B(i)$):}  (If $n_c = 0$, this case does not happen.) This indicates the only way that $P_E(i-1, n_c - 1, n_e)$ contributes to $P_E(i, n_c, n_e)$.
In this case, we must have $B_c(i - 1) = B_c(i) \setminus \{(\mb{a}_i, \mb{y}_i)\}$.
Since $(\mb{a}_i, \mb{y}_i)$ is correct and is not an LC of $B_c(i - 1)$, it has $2^n - 2^{n_c - 1}$ possibilities to occur.
Therefore, Case 3 happens with probability $p (2^n - 2^{n_c - 1})/2^n$.

\emph{Case 4 ($(\mb{a}_i, \mb{y}_i)$ is incorrect and $(\mb{a}_i, \mb{y}_i) \in B(i)$):}  (If $n_e = 0$, this case does not happen.) This indicates the only way that $P_E(i-1, n_c, n_e - 1)$ contributes to $P_E(i, n_c, n_e)$.
In this case, we must have $B_e(i - 1) = B_e(i) \setminus \{(\mb{a}_i, \mb{y}_i)\}$.
Moreover, the error pattern of $(\mb{a}_i, \mb{y}_i)$ cannot be an LC of the error patterns of $B_e(i - 1)$.
Thus, $(\mb{a}_i, \mb{y}_i)$ has $2^{n + l} - 2^{n + n_e - 1}$ possibilities to occur.
Accordingly, Case 4 happens with probability $(1 - p) (2^{n + l} - 2^{n + n_e - 1})/(2^{n+l} - 2^n)$.
At this point, we complete the proof.

\section{Proof of Theorem \ref{theorem: P(G) = sum P_G}}\label{appendix: P(G) = sum P_G}

Recall that $G = (E_1, E_3, F)$.
In \eqref{eqn: P(G) = sum P_G}, $\binom{m}{i} p^i (1-p)^{m-i}$ is the probability that $(\mb{A}, \mb{Y})$ has $i$ correct received symbols.
Given this condition and denoting $\mb{D}$ as the set of the $i$ correct received symbols,  $E_1$ and $E_3$ are independent, and $P_{rk}^*(l, m-i)$ (see Lemma \ref{lemma: rank property no zero row}) is the probability for $E_3$ to happen.
By further considering that $E_3$ happens, the basis $B(\mb{A}, \mb{Y})$ must include all incorrect received symbols such that each incorrect basis element does not attend any LR of $(\mb{A}, \mb{Y}) \setminus B(\mb{A}, \mb{Y})$.
Then, $(E_1, F)$ is equivalent to that the basis of $\mb{D}$ is of size $n$ (i.e., $|B(\mb{D})| = n$) and each basis element of $B(\mb{D})$ attends at least one LR of $\mb{D} \setminus B(\mb{D})$.
Therefore, the probability for $(E_1, F)$ to happen is given by $P_G(i, n, 0)$.
This completes the proof of \eqref{eqn: P(G) = sum P_G}.

We are now to prove \eqref{eqn: P_G}.
Assume $(\mb{A}, \mb{Y})$ only consists of $m$ correct received symbols.
Obviously, we have $P_G(0, 0, 0) = 1$.
For $0 < i \leq m$ and $0 \leq r_0 \leq r \leq n$, we only need to compute how $P_G(i-1, \cdot, \cdot)$ contributes to $P_G(i, r, r_0)$ by taking into account all the situations that happen to the $i$-th received symbol $(\mb{a}_i, \mb{y}_i)$.
Since random fountain codes are considered, $(\mb{a}_i, \mb{y}_i)$ has $2^n$ equally like possibilities to occur.
Moreover, recall that $B(i), i \in [m]$ denotes the basis of $(\mb{A}[i], \mb{Y}[i])$  and $B(0) = \emptyset$.
According to the third assumption at the beginning of Section \ref{section: performance analysis}, $(\mb{a}_i, \mb{y}_i) \in B(i)$ iff $(\mb{a}_i, \mb{y}_i)$ is not an LC of $B(i-1)$.

\emph{Case 1 ($(\mb{a}_i, \mb{y}_i) \in B(i)$):}
(If $r < 0$ or $r_0 < 0$, this case does not happen.)
This indicates the only way that $P_G(i-1, r - 1, r_0 - 1)$ contributes to $P_G(i, r, r_0)$.
In this case, we must have that $(\mb{a}_i, \mb{y}_i)$ is not an LC of $B(i-1)$ and $|B(i-1)| = r-1$.
Thus, $(\mb{a}_i, \mb{y}_i)$ has $2^n - 2^{r - 1}$ possibilities to occur, indicating that $P_G(i-1, r - 1, r_0 - 1)$ contributes to $P_G(i, r, r_0)$ with probability $1 - 2^{r-1-n}$.

\emph{Case 2 ($(\mb{a}_i, \mb{y}_i) \notin B(i)$):} This indicates the only way that $P_G(i-1, r, r'_0)$ contributes to $P_G(i, r, r_0)$ for $r_0 \leq r'_0 \leq r$.
In this case, $(\mb{a}_i, \mb{y}_i)$ is an LC of $B(i-1)$ and we have $B(i) = B(i-1)$.
Denote $B_0$ as the set of the $r'_0$ basis elements of $B(i-1)$ that do not attend the LRs of $(\mb{A}[i-1], \mb{Y}[i-1]) \setminus B(i-1)$.
Then, there are exactly $r_0$ basis elements of $B_0$ that do not attend the LR of $(\mb{a}_i, \mb{y}_i)$.
As a result, $(\mb{a}_i, \mb{y}_i)$ has $\binom{r'_0}{r_0} 2^{r - r'_0}$ possibilities to occur, indicating that $P_G(i-1, r, r'_0)$ contributes to $P_G(i, r, r_0)$ with probability $\binom{r'_0}{r_0} 2^{r - r'_0 - n}$.
At this point, we complete the proof.

\section{Proof of Theorem \ref{theorem: probability for attending LR}}\label{appendix: probability for attending LR}

Use the notations of Theorem \ref{theorem: probability for attending LR}.
Denote $B_c$ and $B_e$ as the sets of correct and incorrect received symbols of $B_r$, respectively.
We have $|B_c| = n$, $0 \leq |B_e| = r - n  \leq l$, and $B_c \cup B_e = B_r$.
All the LCs of $B_c$ result in $V(\mb{X})$ (see \eqref{eqn: V(x)}), which is the set of all possible encoded symbols/correct received symbols.
Denote $L_C$ as the set of all the LCs of $B_r$.

Given that $(\mb{a}, \mb{y}) \in L_C$,  each received symbol of $B_c$ attends the LR of $(\mb{a}, \mb{y})$ with probability $p_c = 1/2$.
On the other hand, only when $(\mb{a}, \mb{y})$ is incorrect (i.e., $(\mb{a}, \mb{y}) \in L_C \setminus V(\mb{X})$), a received symbol of $B_e$ can have a chance to attend the LR of $(\mb{a}, \mb{y})$.
$(\mb{a}, \mb{y})$ is incorrect with probability
\begin{align*}
    &\mbb{P}\big((\mb{a}, \mb{y}) \in L_C \setminus V(\mb{X}) \mid (\mb{a}, \mb{y}) \in L_C \big)
    \\=& \frac{\mbb{P}\big((\mb{a}, \mb{y}) \in L_C \setminus V(\mb{X})\big)}{\mbb{P}\big((\mb{a}, \mb{y}) \in L_C \big)}
    \\=& \frac{(1 - p) (2^{r-n} - 1)/(2^{l} - 1)}{p + (1 - p) (2^{r-n} - 1)/(2^{l} - 1)}.
\end{align*}
Given $(\mb{a}, \mb{y}) \in L_C \setminus V(\mb{X})$, each received symbol of $B_e$ attends the LR of $(\mb{a}, \mb{y})$ with probability
\[
    \frac{|L_C| / 2}{|L_C \setminus V(\mb{X})|} = \frac{2^{r - n - 1}}{2^{r - n} - 1}.
\]
Thus, we have
\begin{align*}
    p_e =& \frac{(1 - p) (2^{r-n} - 1)/(2^{l} - 1)}{p + (1 - p) (2^{r-n} - 1)/(2^{l} - 1)} \frac{2^{r - n - 1}}{2^{r - n} - 1}
    \\=& \frac{(1 - p) 2^{r-n-1}/(2^{l} - 1)}{p + (1 - p) (2^{r-n} - 1)/(2^{l} - 1)}
    \\\leq& (1 - p) 2^{l - 1}/(2^{l} - 1)
    \\< &p_c,
\end{align*}
where the last inequality is due to \eqref{eqn: p max}.


\ifCLASSOPTIONcaptionsoff
  \newpage
\fi

\bibliographystyle{IEEEtran}
\bibliography{myreference}

\end{document}